# The intermittent vibration observed by Reynolds in a cylindrical tube - Explanation of Reynolds' experiment

Qian Zuwen

Institute of Acoustics, Chinese Academy of Sciences, Beijing 100190, China

**Abstract**. In a cylindrical tube filled with incompressible fluid, the variable parameter method is applied to solve the vorticity motion equation, and the obtained results are substituted into the Poisson equation satisfied by the energy density. The static distribution of energy density in the tube and the energy density of the flowing fluid after a certain spatial sampling interval are obtained. Further numerical calculations are conducted on the energy density of the flowing fluid, the results show that the intermittent vibration observed by Reynolds 140 years ago can only be obtained with the appropriate spatial sampling interval. The energy density in the tube is proportional to $(10^6)^N$ to $(10^7)^N$, where N is an integer  It  tends to infinity with increasing N.

**Introduction**

Since the publication of Reynolds' classic experiment [1] 140 years ago, researchers have continually strived to comprehend the formation mechanism of turbulence. As the introduction of this paper, we will commence by citing several authors' depictions for the research status of turbulent flow in tubes.

(1) "Ever since the pioneering experimental work of Osborne Reynolds (1883), the issue of how and why the fluid flow along a circular pipe changes from being laminar (highly ordered in space and time) to turbulent (highly disordered in both space and time) as the flow rate increases has intrigued physicists, mathematicians and engineers alike. The problem, of course, is not that we do not know the governing equations of motion since we do—they are the celebrated Navier–Stokes equations. Rather the challenge is to extract the relevant information from this notoriously difficult set of nonlinear equations to rationalize what we see."[2];

(2)"Further elucidation, confirmation, explaining existing experimental observations,

and useful application of the insights into this exciting mechanism of transition to turbulence are the challenge now"[3];

(3) "One circumstance that complicates this problem is that laminar pipe flow is stable to infinitesimal perturbations, and therefore in order to trigger turbulence, a disturbance of finite amplitude is required. What makes matters even more difficult is that at low Re, turbulence is transient. Here, turbulence occurs in the form of localized patches called puffs that are embedded in the surrounding laminar flow and decay according to a memoryless process (that is, independent of their previous history)."[4]

The results obtained from the Reynolds experiment have intrigued researchers in fluid mechanics, presenting several mysteries. Among these, the two most notable issues are [1]:

(i) "Another phenomenon very marked in the smaller tubes, was the intermittent character of the disturbance. The disturbance would suddenly come on through a certain length of the tube and pass away and then come on again, giving the appearance of flashes, and these flashes would often commence successively at one point in the pipe";

(ii) "Under no circumstances would the disturbance occur nearer to the trumpet than about 30 diameters in any of the pipes, and the flashes generally, but not always, commenced at about this distance".

To the best of the author's knowledge, these two issues have not been effectively resolved to date.

In reference [5], the author of this article computed the energy density within a cylindrical tube, and the findings revealed that the energy density in the tube increases gradually due to axial flow and viscous traction. As the Reynolds number increased, the energy density in the tube eventually approaches infinity. This paper serves as a summary of references [5-8], upon which further modifications and supplements have been made.

**2. Theory**

We begin with the vortex Equations (cf. Appendix 1)

$$\frac{\partial \mathbf{\Omega}}{\partial t} - \nu \nabla^2 \mathbf{\Omega} = -(\mathbf{u} \cdot \nabla)\mathbf{\Omega} \qquad (1)$$

and Poison equation

$$\nabla^2 (\frac{1}{2}u^2 + \frac{P}{\rho}) = (\nabla \times u)^2 + u \cdot \nabla^2 u \tag{2}$$

The initial flow rate in the tube is

$$\begin{aligned} \mathbf{u}^{(0)} &= \mathbf{i}_r u_r^{(0)} + \mathbf{i}_z u_z^{(0)} \\ u_z^{(0)} &= V_0 (1 - \frac{r^2}{R^2}) \\ u_r^{(0)} &= 0 \end{aligned} \tag{3}$$

At first, dimensionless equation (1) will be executed. Let

$$\begin{aligned} \frac{V_0}{R} t &= t^*, \quad \frac{r}{R} = r^*, \quad \frac{z}{R} = z^*, \quad u/V_0 = u^* \\ \mathbf{\Omega} &= \nabla \times \mathbf{u} = \frac{V_0}{R} \mathbf{\Omega}^*, \quad \nabla^2 = \frac{1}{R^2} \nabla^{*2} \\ u_z^{*(0)} &= (1 - r^{*2}) \\ u_r^{*(0)} &= 0 \end{aligned} \tag{4}$$

and the equation (1) becomes

$$\frac{\partial \mathbf{\Omega}^*}{\partial t^*} - \frac{1}{\text{Re}} \nabla^{*2} \mathbf{\Omega}^* = -\mathbf{u}^* \cdot \nabla^* \mathbf{\Omega}^* \tag{5}$$

By Laplace transformation of equation (5), one has

$$\begin{aligned} \nabla_r^{*2} \bar{\mathbf{\Omega}}^* + \kappa^2 \bar{\mathbf{\Omega}}^* &= \text{Re}\, L[\mathbf{u}^* \cdot \nabla^* \mathbf{\Omega}^*] \\ \bar{\mathbf{\Omega}}^* &= \int_0^\infty \mathbf{\Omega}^* e^{-\gamma t^*} dt^* \end{aligned} \tag{6}$$

where

$$\kappa = \sqrt{\alpha^2 - \text{Re}\, \gamma} \tag{7}$$

and $\alpha$ is the attenuation coefficient, Re is the Reynolds number, and $L[F]$ is the Laplace transformation of F. (for convenience, the asterisk will be omitted from now on).

When the first expression in the right end of equation (6) is equal to zero, its homogeneous equation is obtained i.e.

$$\nabla_r^2 \mathbf{\Omega}^{(0)} + \kappa^2 \mathbf{\Omega}^{(0)} = 0$$

By Laplace transformation of the above formula yields

$$\nabla_r^2 \bar{\mathbf{\Omega}}^{(0)} + \kappa^2 \bar{\mathbf{\Omega}}^{(0)} = 0$$

According to the method of separating variables, the solution of the homogeneous equation (6)

is

$$\bar{\mathbf{\Omega}}^{(0)} = A^{(0)} J_0(\kappa Rr)e^{-\alpha Rz} \tag{9}$$

$A^{(0)}$ is a constant. If the right end is not equal to zero, then

$$\nabla_r^2 \bar{\mathbf{\Omega}} + \kappa^2 \bar{\mathbf{\Omega}} = \operatorname{Re} L[\mathbf{u} \cdot \nabla \mathbf{\Omega}] \tag{10}$$

On the right end of the above equation, the homogeneous solution $\bar{\mathbf{\Omega}}^{(0)}$ is used to represent $\mathbf{\Omega}$ approximately and the velocity component in the z-direction $\mathbf{i}_z u_z^{(0)}$ is used to denote $\mathbf{u}$, therefore, the above equation becomes

$$\nabla_r^2 [\bar{\mathbf{\Omega}}^{(1)}] + \kappa^2 [\bar{\mathbf{\Omega}}^{(1)}] = \frac{1}{\alpha R} \{u_r^{(0)} \frac{\partial \mathbf{\Omega}^{(0)}}{\partial r} + u_z^{(0)} \frac{\partial \mathbf{\Omega}^{(0)}}{\partial z}\} \tag{11}$$

Substituting $\mathbf{u}^{(0)}$ and $\mathbf{\Omega}^{(0)}$ into the equation (10) yields

$$\nabla_r^2 \bar{\mathbf{\Omega}}^{(1)} + \kappa^2 \bar{\mathbf{\Omega}}^{(1)} = \varepsilon(1 - r^2)\bar{\mathbf{\Omega}}^{(0)} \tag{12}$$

$$\operatorname{Re} \alpha R = \varepsilon \tag{13}$$

## 3. The solution of equation (12)

Using the variable parameter method (cf. Appendix 2) to find the special solution of equation (12), that is

$$\bar{\Omega}^{(1)}(r, z; \gamma) = A^{(1)}(r; \gamma) J_0(\kappa Rr) e^{-\alpha Rz}$$

(14)

Substituting (14) into equations (12) to get

$$\left.\begin{array}{l} A_r^{(1)}(r;\gamma) = 0 \\ A_z^{(1)}(r;\gamma) = \varepsilon A^{(0)} f^{(1)}(r) \\ f^{(1)}(r) = \int \frac{dr}{r[J_0(\kappa Rr)]^2} \int r(1-r^2) J_0(\kappa Rr) J_0(\kappa Rr) dr \end{array}\right\} \tag{15}$$

The inverse transform of (14) is

$$\Omega^{(1)}(r, z) = \varepsilon A^{(0)} f^{(1)}(r) J_0(\alpha Rr) e^{-\alpha Rz}$$

(16)

It can be similarly obtained following results from Appendix 3

$$u_z^{(1)}(r,z) = 1 - \int_0^r A^{(1)}(r) J_0(\kappa Rr) e^{-\alpha Rz} dr$$

$$u_r^{(1)}(r,z) = 0$$

$$\nabla^2 u_r^{(1)} = -\alpha A^{(1)}(r,0) J_0(\alpha Rr) e^{-\alpha Rz}$$

$$\nabla^2 u_z^{(1)} = -(\frac{\partial}{\partial r} + \frac{1}{r}) A^{(1)}(r) J_0(\kappa Rr) e^{-\alpha Rz}$$

(17)

## 4. Solutions of equation (2)

For flow in a tube, because the space is bounded, the solution of the Poisson equation (2) consists of two volume integrals and a surface integral, that is

$$(\frac{1}{2}u^2 + \frac{P}{\rho}) = D(r,z) + G(r,z) + S(r,z) \tag{18}$$

where

$$D(r,z) = \int_{\tau_s} g(r,z;r_s,z_s)[\Omega]^2 d\tau_s, \quad G(t;r,z) = \int_{\tau_s} g(r,z;r_s,z_s) \boldsymbol{u} \cdot \nabla^2 \boldsymbol{u} d\tau_s,$$

$$S(r,z) = \int_s \{(\frac{1}{2}u^2(R,Z) + \frac{P(R,Z)}{\rho}) \frac{\partial g(r,z;R,Z)}{\partial R} -$$

$$g(r,z;R,Z) \frac{\partial[(\frac{1}{2}u^2(R,Z) + \frac{P(R,Z)}{\rho})]}{\partial R}\} 2\pi R dZ \tag{19}$$

$$= \int_s \{\frac{P(R,Z)}{\rho} \frac{\partial g(r,z;R,Z)}{\partial R} - g(r,z;R,Z) \frac{\partial[(\frac{P(R,Z)}{\rho})]}{\partial R}\} 2\pi R dZ$$

## 5. Integrals $D$ and $G$

### 5.1. Integral $D$

Substituting equation (16) into the first equation of equation (19) yields

$$D = \int_0^1 \int_0^\infty \frac{[\Omega(r_s,z_s)]^2}{|\mathbf{r}-\mathbf{r}_s|} 2\pi r_s dz_s dr_s$$

$$= 2\pi e^{-2\alpha Rz} A_0^2 \varepsilon^2 \int_0^1 [f^{(1)}(r_s)]^2 J_0^2(\alpha Rr_s) r_s [\int_0^\infty I(r_s,z_s) dz_s] dr_s, \tag{20}$$

where

$$I(r_s,z_s) = \frac{e^{-2\alpha Rz_s}}{\sqrt{(r_s-r)^2 + (z_s-z)^2}}. \tag{21}$$

$$D = \int_0^1 \int_0^l \frac{\Omega(r_s, z_s)^2}{4\pi |\mathbf{r} - \mathbf{r}_s|} 2\pi r_s dz_s dr_s = \int_0^1 B_1(r_s)[\int_0^l I(r_s, z_s; r, z) dz_s] r_s dr_s \quad (22)$$

$$B_1(r_s) = \frac{1}{2}[\varepsilon A^{(0)} f^{(1)}(r_s)]^2 J_0^2(\alpha R r_s) \quad (23)$$

where the integral limit is (0, *l*) instead of (0, ∞) with *l* being the length of the tube.

## 5.2 Integral G

Put (17) into the second expression of equation (19) to get (cf. Appendix 4)

$$G(r, z; t) = \int_{\tau_s} \boldsymbol{u} \cdot \nabla^2 \boldsymbol{u} \ g(r, z; r_s, z_s) d\tau_s$$

$$= \int \frac{\boldsymbol{u}(r_s, z_s) \cdot \nabla^2 \boldsymbol{u}(r_s, z_s)}{4\pi \sqrt{(r_s - r)^2 + (z_s - z)^2}} 2\pi r_s dr_s dz_s \quad (24)$$

$$= \int_0^1 B_2(r_s)[\int_0^l I(r_s, z_s; r, z) dz_s] r_s dr_s$$

$$B_2(r) = -\frac{1}{2}(\varepsilon A^{(0)})^2 \{(\frac{\partial}{\partial r} + \frac{1}{r})[f^{(1)}(r) J_0(\alpha R r)]\} \{1 - \int_0^r A^{(1)}(r) J_0(\kappa R r) e^{-\alpha R z} dr\}. \quad (25)$$

The contribution of the surface integral only affects the fluid near the wall, and has little effect on the fluid in other parts, so it can be ignored.

## 5.3 Integrals of energy density

Substituting (20)-(25) into (18) to get

$$E(r, z) = D + G \quad (26)$$

then one has

$$E(r, z) = \int_0^1 B(r_s)[\int_0^l I(r_s, z_s; r, z) dz_s] r_s dr_s \quad (27)$$

where

$$B(r_s) = B_1(r_s) + B_2(r_s) \quad (28)$$

From (A2.21), (A3.13) and (A4.13), one has

$$f^{(1)}(x) \approx \frac{1}{4} x^2 (1 - \frac{1}{24} x^2) \quad (29)$$

$$\varepsilon A^{(0)} = \frac{1}{\int_0^1 f^{(1)}(r) J_0(\alpha R r) dr} \quad (30)$$

and

$$B(r) = B_1(r) + B_2(r) = (\varepsilon A^{(0)})^2 \{\frac{3}{32}r^4 - \frac{343}{5760} + \frac{1345}{129024} - \frac{109}{184320} + \frac{1}{98304}\} \quad (31)$$

respectively. Now we will divide (27) into four equations, i.e.

$$E(r,z) = \int_0^1 B(r_s)[\int_0^l I(r_s, z_s; r, z)dz_s]r_s dr_s = \{\int_0^r + \int_r^1\}\{\int_0^z + \int_z^l\} =$$

$$= \int_0^r \int_0^z + \int_0^r \int_z^l + \int_r^1 \int_0^z + \int_r^1 \int_z^l = I + II + III + IV \quad (32)$$

At first, we start to calculate integral I in equation (32),

$$I = \int_0^r B(r_s)r_s \int_0^z \frac{e^{-2\alpha R z_s}}{\sqrt{(r_s - r)^2 + (z_s - z)^2}} dz_s dr_s$$

Obviously, in the above integration one has

$$z_s \leq z, \quad r_s \leq r$$

and we can make following transform

$$Z = z - z_s = (r - r_s)\tan\psi$$

to get

$$I = \int_0^r B(r_s)r_s \int_0^z \frac{e^{-2\alpha R z_s}}{\sqrt{(r_s - r)^2 + (z_s - z)^2}} dz_s dr_s$$

$$= e^{-2\alpha R z} \int_0^r B(r_s)r_s \int_{z_s=0}^{z_s=z} \frac{e^{2\alpha R Z}}{\sqrt{(r_s - r)^2 + Z^2}} d(-Z) dr_s \quad (33)$$

$$= e^{-2\alpha R z} \int_0^r B(r_s)r_s \int_0^{\psi_1} \frac{e^{2\alpha R(r-r_s)\tan\psi}}{\cos\psi} d\psi dr_s$$

where

$$\psi_1 = \arctan(\frac{z}{r - r_s}) \quad (34)$$

Similarly, due to

$$z_s \geq z, \quad r_s \leq r, \quad z_s - z = Z \geq 0$$

in the second integral II, one has

$$II = e^{-2\alpha R z} \int_0^r B(r_s)r_s \int_z^l \frac{e^{-2\alpha R(z_s - z)}}{\sqrt{(r_s - r)^2 + (z_s - z)^2}} dz_s dr_s$$

we can take following transform

$$Z=(r-r_s)\tan\psi$$

to get

$$\text{II} = e^{-2\alpha Rz}\int_0^r B(r_s)r_s\int_z^l \frac{e^{-2\alpha R(z_s-z)}}{\sqrt{(r_s-r)^2+(z_s-z)^2}}dz_s dr_s$$

$$= e^{-2\alpha Rz}\int_0^r B(r_s)r_{ss}\int_0^{l-z} \frac{e^{-2\alpha RZ}}{\sqrt{(r_s-r)^2+Z^2}}dZdr = e^{-2\alpha Rz}\int_0^r B(r_s)r_s\int_0^{\psi_2} \frac{e^{-2\alpha R|r-r_s|\tan\psi}}{\cos\psi}d\psi dr_s \quad (35)$$

where

$$\tan\psi_2 = \frac{l-z}{|r_s-r|} \quad (36)$$

In the third integral

$$\text{III} = \int_r^1 B(r_s)r_s\int_0^z \frac{e^{-2\alpha Rz_s}}{\sqrt{(r_s-r)^2+(z_s-z)^2}}dz_s dr_s$$

where

$$z_s \leq z,\ r_s \geq r,\ z-z_s = Z \geq 0$$

we execute the transform

$$Z=(r_s-r)\tan\psi$$

Substituting it into III yields

$$\text{III} = \int_r^1 B(r_s)r_s\int_0^z \frac{e^{-2\alpha Rz_s}}{\sqrt{(r_s-r)^2+(z_s-z)^2}}dz_s dr_s$$

$$= e^{-2\alpha Rz}\int_r^1 B(r_s)r_s\int_{z_s=0}^{z_s=z} \frac{e^{-2\alpha R(-Z)}}{\sqrt{(r_s-r)^2+Z^2}}d(-Z)dr_s \quad (37)$$

$$= e^{-2\alpha Rz}\int_r^1 B(r_s)r_s\int_0^{\psi_1} \frac{e^{2\alpha R|r_s-r|\tan\psi}}{\cos\psi}d\psi dr_s$$

For the integral IV,

$$\text{IV} = \int_r^1 B(r_s)r_s\int_z^l I(r_s,z_s;r,z)dz_s dr_s$$

where

$$z_s \geq z,\ r_s \geq r,\ z_s-z = Z > 0.$$

Take a transform

$$Z=(r_s - r)\tan\psi$$

By substitution of it into IV, one has

$$\begin{aligned}
\text{IV} &= \int_r^1 B(r_s)r_s \int_z^l \frac{e^{-2\alpha R z_s}}{\sqrt{(r_s-r)^2+(z_s-z)^2}} dz_s dr_s \\
&= e^{-2\alpha R z}\int_r^1 B(r_s)r_s \int_z^l \frac{e^{-2\alpha R(z_s-z)}}{\sqrt{(r_s-r)^2+(z_s-z)^2}} dz_s dr_s \quad (38) \\
&= e^{-2\alpha R z}\int_r^1 B(r_s)r_s \int_0^{\psi_2} \frac{e^{-2\alpha R|r_s-r|\tan\psi}}{\cos\psi} d\psi dr_s
\end{aligned}$$

Adding four integrals to get

$$\begin{aligned}
E(r,z)&=\text{I}+\text{III}+\text{II}+\text{IV}=e^{-2\alpha R z}\int_0^1 B(r_s)r_s \int_0^{\psi_1}\frac{e^{2\alpha R|r_s-r|\tan\psi}}{\cos\psi}d\psi dr_s \\
&+ e^{-2\alpha R z}\int_0^1 B(r_s)r_s \int_0^{\psi_2}\frac{e^{-2\alpha R|r_s-r|\tan\psi}}{\cos\psi}d\psi dr_s
\end{aligned} \quad (39)$$

## 6. Energy density integral in a cylindrical tube

Perform multiple partial integrals on the equation (39) finally obtain (cf. Appendix 5)

$$E(r,z)=e^{-2\alpha R z}\sum_{n=1}^{N}\int_0^1 B(r_s)g_n(r_s-r,z)r_s dr_s + e^{-2\alpha R z}\int_0^1 B(r_s)Y_N(r_s-r,z)r_s dr_s \quad (40)$$

where

$$\left.\begin{aligned}
g_1(r_s-r,z) &\approx e^{-2\alpha R z}\frac{1}{2\alpha R|r_s-r|}[e^{2\alpha R z}\cos\psi_1 - e^{-2\alpha R(l-z)}\cos\psi_2] \\
g_2(r_s-r,z) &\approx \frac{1}{(2\alpha R|r_s-r|)^2}[\cos^2\psi_1 \sin\psi_1 + \cos^2\psi_2 \sin\psi_2 e^{-2\alpha R l}] \\
g_3(r_s-r,z) &= -\frac{1}{(2\alpha R|r_s-r|)^3}\{[\cos^2\psi_1 \frac{d}{d\psi_1}\cos^2\psi_1 \frac{d}{d\psi_1}(-\cos\psi_1) - \\
&\quad -\cos^2\psi_2 \frac{d}{d\psi_2}\cos^2\psi_2 \frac{d}{d\psi_2}(-\cos\psi_2)e^{2\alpha R l}\} \\
g_4(r_s-r,z) &= \frac{1}{(2\alpha R|r_s-r|)^4}e^{-2\alpha R z}[(\cos^2\psi_1 \frac{d}{d\psi_1})^3(-\cos\psi_1)e^{2\alpha R z} \\
&\quad + (\cos^2\psi_2 \frac{d}{d\psi_2})^3(-\cos\psi_2)e^{-2\alpha R(l-z)}]
\end{aligned}\right\} \quad (41)$$

$$g_n(r_s - r, z) = \frac{1}{(2\alpha R |r_s - r|)^n} e^{-2\alpha Rz}[(-\cos^2 \psi_1 \frac{d}{d\psi_1})^{n-1}(\cos \psi_1)e^{2\alpha Rz}$$
$$+ (-1)^n (-\cos^2 \psi_2 \frac{d}{d\psi_2})^{n-1}(\cos \psi_2)e^{2\alpha R(l-z)}]$$
(42)

$$Y_n(r_s - r, z) = -\frac{1}{(2\alpha R|r_s-r|)^n} e^{-2\alpha Rz} \int_0^{\psi_1} e^{2\alpha R|r_s-r|\tan\psi}(-1)^n(-\frac{d}{d\psi}\cos^2\psi)^{n-1}[\frac{d}{d\psi}(\cos\psi)]d\psi +$$
$$- \frac{1}{(2\alpha R|r_s-r|)^n} e^{-2\alpha Rz} \int_0^{\psi_2} e^{2\alpha R|r_s-r|\tan\psi}(-1)^n(-\frac{d}{d\psi}\cos^2\psi)^{n-1}[\frac{d}{d\psi}(\cos\psi)]d\psi$$

and

$$\tan\psi_1 = \frac{z}{|r_s - r|}, \quad \tan\psi_2 = \frac{l-z}{|r_s - r|}, \quad \cos\psi_1 = \frac{1}{\sec\psi_1} = \frac{|r_s - r|}{\sqrt{|r_s - r|^2 + z^2}},$$
$$\sin\psi_1 = \sqrt{1 - \cos^2\psi_1} = \sqrt{1 - \frac{|r_s - r|^2}{|r_s - r|^2 + z^2}} = \sqrt{\frac{z^2}{|r_s - r|^2 + z^2}}$$
(43)

Rewrite (40) as

$$E_s(r,z) = E(r,z) - e^{-2\alpha Rz}\int_0^1 B(r_s)Y_N(r_s - r, z)r_s dr_s = e^{-2\alpha Rz}\sum_{n=1}^{N}\int_0^1 B(r_s)g_n(r_s - r, z)r_s dr_s \quad (44)$$

Substituting (31) and (41) into (44) and executing numerical computation yield results regarding the relationship between energy density and $z$ (see Figures 1(*a*), 1(*b*), 1(*c*) and 1(*d*)) for $N$=1-4, where the horizontal axes in the figures represent the axial variable $z$, and the vertical axes represent $E_s(r,z)$.

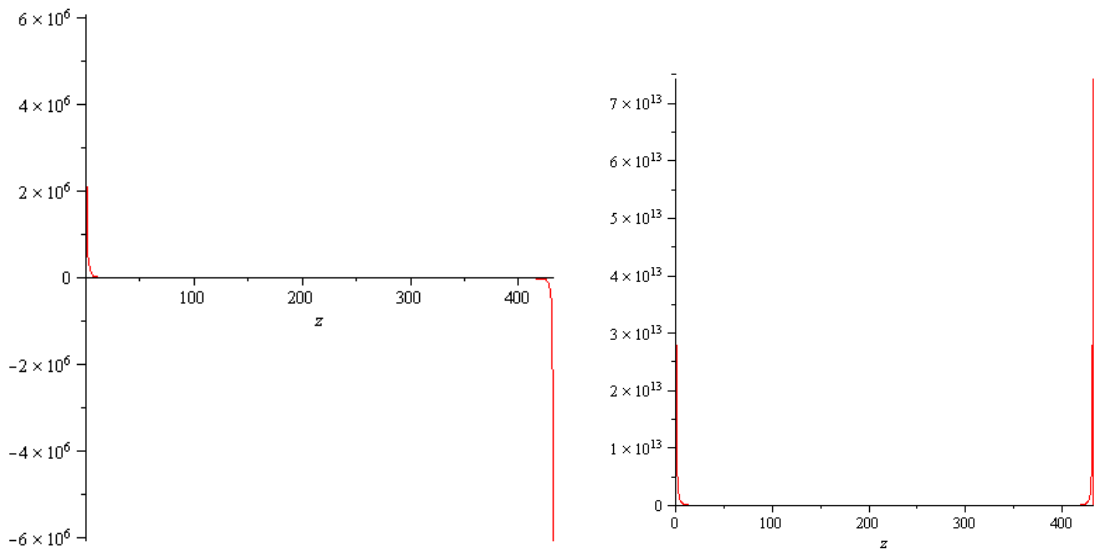

Figure 1(*a*). *N*=1        Figure 1(*b*). *N*=2

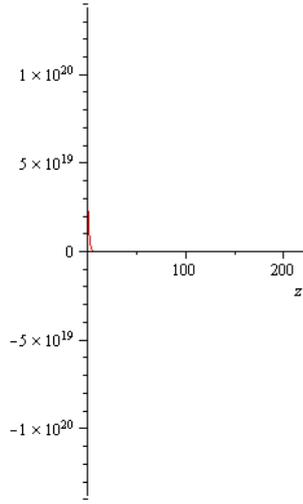 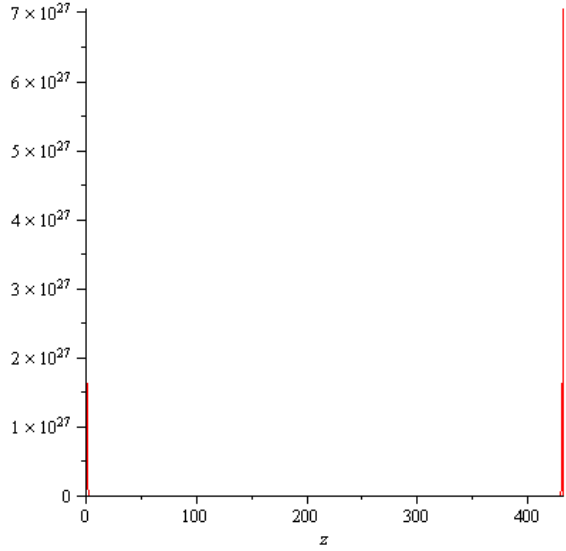

1(*c*). N= 3        1(*d*). N=4

Figure 1.  the relationship between energy density and *z* for N=1, 2, 3 and 4.

From the energy density distribution diagram shown in Figures 1-4, it can be observed that when N=1 there are only spikes at z=0 and z=*l*, while the rest are close to zero, particularly, the spike at z=*l* is negative; For N=2 and N=4, there are positive spikes with different values at z=0 and z=*l*, while other parts are close to zero; For N=3 the energy density distribution is similar to N=1, except for a small spike at z=*l*.

## 7. Energy density of flowing fluids

Figure 1 illustrates the distribution of energy density along the z-direction within a cylindrical tube, referred to as the static distribution of energy density. When a fluid with a speed $\bar{u}_{0z}$ moves from left to right for a period of time $\Delta t$, of which the total energy density of the fluid $E_m(r,z)$ should be the summation of the energy density $E_s(r,z)$ that brought from its previous location $(\Delta z = 0)$ and the static energy density $E_s(r, z + \Delta z)$, $\Delta z = \bar{u}_{0z}\Delta t$, at the new location where the fluid move on i.e.

$$E_m(r,z;1) = E_s(r,z) + E_s(r,z+\Delta z) \qquad (45)$$

At this point, the energy density of the fluid is the sum of the energy densities in the two images before and after moving a spatial interval $\Delta z$. If the fluid continues to move forward for another distance $\Delta z$, the energy density field of the fluid becomes

$$E_m(r,z;2) = E_s(r,z) + E_s(r,z+\Delta z) + E_s(r,z+2\Delta z)$$

If the fluid continues to flow forward for K-th distance, the energy density field of the flowing fluid becomes

$$E_m(r,z;K) = \sum_{k=0}^{K} E_s(r,z+k\Delta z) = E_m(r,z;K=0) + E_m(r,z;K>0) \qquad (46)$$

where

$$E_m(r,z;K=0) = E_s(r,z) \qquad (47)$$

and

$$E_m(r,z;K>0) = \sum_{k=1}^{K} E_s(r,z+k\Delta z) \qquad (48)$$

At this time, the fluid moves from left to right, which is equivalent to a static image of the energy density moving from right to left. In the following, it will be seen that the dominant contribution is only the spike at the point $z = l$ in figures 1-4, while the energy density of the other parts is almost zero. When the energy density value at $z = l$ is very small, only weak intermittent vibration can occur. If the energy density is equal to zero, the sum of the results will not result in intermittent vibration.

Substitute equations (41)-(44) into equation (46) and select $\Delta z$, $k = 0,1,2,\cdots,K$ for trial calculation of energy density of the fluid. For example, substituting $\Delta z = 9.55$ into equation (46) yields figures 2(*a*), 2(*b*), 2(*c*) and 2(*d*) for N=2. Figure 2(*e*) is a schematic diagram of Reynolds' experimental results. By the way, Reynolds' paper did not explicitly state how many intermittent oscillations he observed, only showed results in Fig.16 of the paper. However, the numerical results obtained in both cases above-mentioned agree with the results observed in the Reynolds' experiment. Figures 3 and 4 denote the computation results for *N*=3 and N=4, respectively. These results indicate that intermittent vibrations occur outside the range of 60 times the radius, which is consistent with the results observed in Reynolds' experiments.

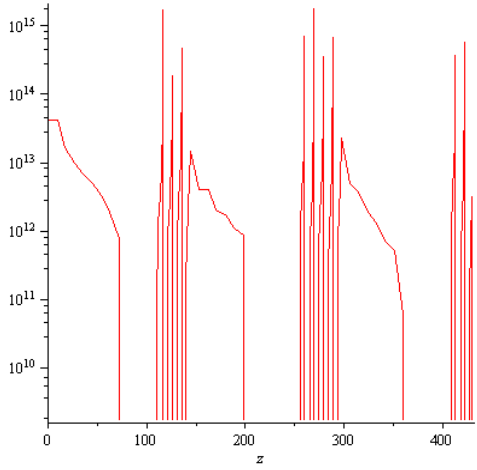 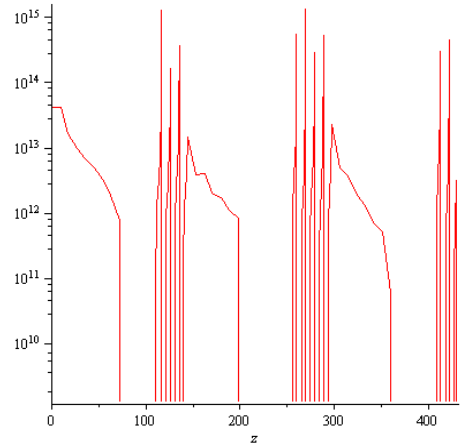

2(*a*) N=2, x=0.75, Δ*z* =9.55, *K*=45     2(*b*) N=2, x=0.7, Δ*z* =9.55, *K*=45

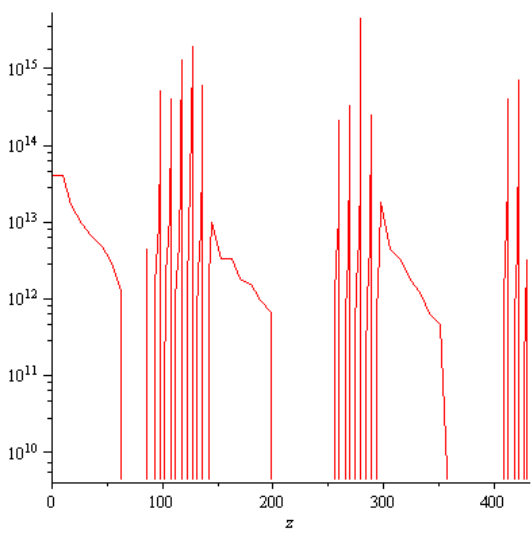 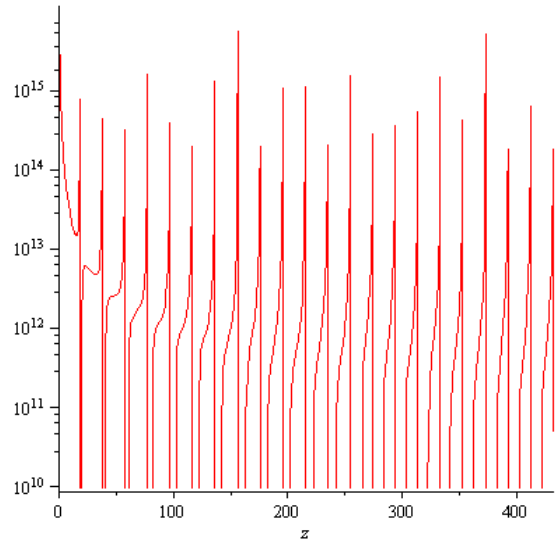

2(*c*) N=2, x=0.85, Δ*z* =9.54, *K*=45     2(*d*) N=2, x=0.85, Δ*z* =19.7, *K*=45

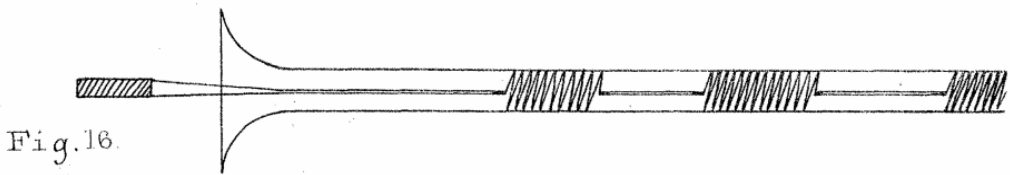

2(*e*) Reynolds' experimental results (From Figure 16 of Ref.[1] )

**Figure 2. Energy density of the fluid for *N*=2**

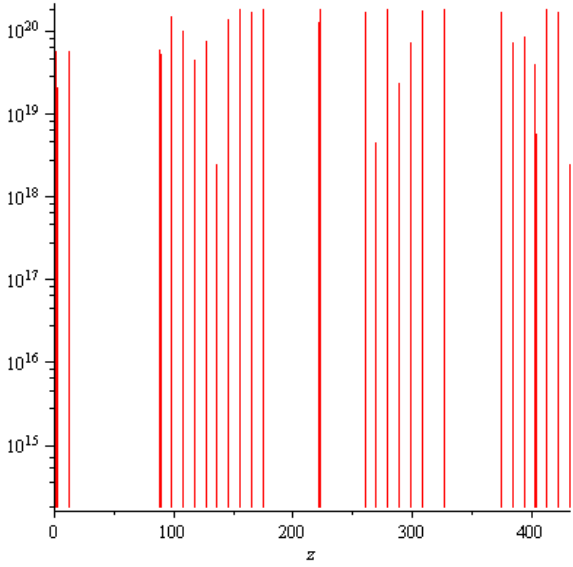
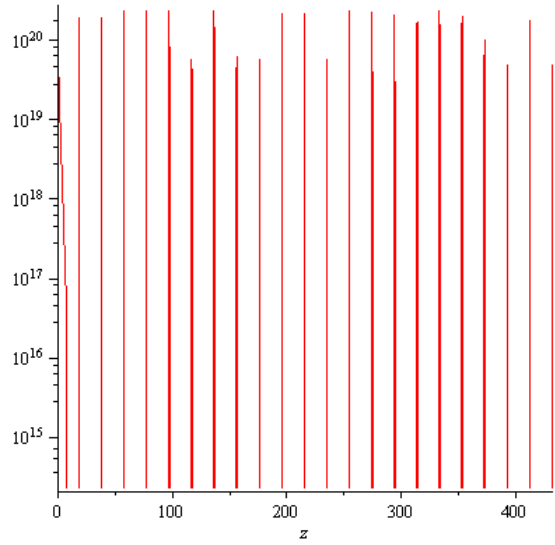

(*a*)N=3, *x*=0.25, Δ*z* =9.535, *K*=45     3(*b*) N=3, *x*=0.15, Δ*z* =19.7, *K*=45

**Figure 3. Energy density of the fluid for *N*=3**

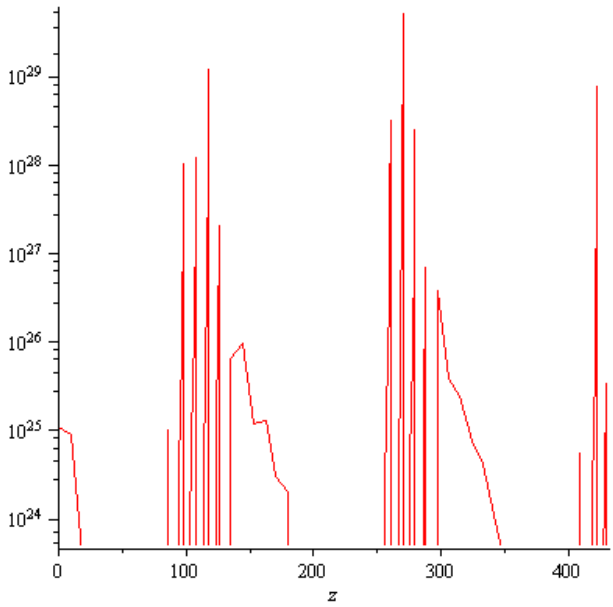
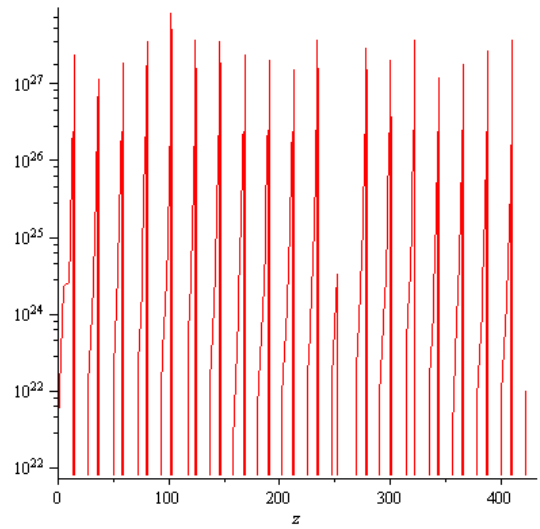

4 (*a*) N=4, Δ*z* =9.53, *x*=0.5, ,*K*=45     4 (*b*) N=4, Δ*z* =22, *x*=0.25, *K*=19

**Figure 4. Energy density of the fluid for *N*=4**

**Discussion**

According to the Fourier series expansion theorem, a function that satisfies the Dirichlet condition within the interval (0, *l*) can be expanded into a Fourier series to represent it. Let's now study the case where the relationship between amplitude and z is a sine function, i.e.

$$E(z,t) = E(t)\sin(\frac{2\pi}{d} z) \tag{49}$$

where d is the wavelength. If the tube length is equal to 1/4 wavelength, i.e

$$l = \frac{d}{4} \tag{50}$$

which indicates that at $z = 0$ where is a node and at $z = l$ where is an antinode, and we refer to this normal mode as a 1/4 wavelength oscillator; If the length of the tube is equal to 3/4 of the wavelength, i.e

$$l = \frac{3d}{4} = \frac{d}{2} + \frac{d}{4} \tag{51}$$

then the normal mode consists of one standing wave unit and one 1/4 wavelength oscillator; If

$$l = \frac{5d}{4} = \frac{d}{2} + \frac{d}{2} + \frac{d}{4} \tag{52}$$

the normal mode consists of two standing wave units and one 1/4 wavelength oscillator; and so on.

Now we will use above discussions to analyze the numerical computation results. Obviously, figures 1(*a*)-1(*d*) respectively represent the energy density $E_m(r,z;K)$ for N=1-4, the images of them are almost Dirac function distributed in $z = 0$ and $z = l$, while in other places it is almost zero. According to the above discussion, the presence or absence of standing wave images depends on the presence or absence of positive spikes at $z = l$. For N=1, there is no positive spike in the energy density, so there is no standing wave image (or intermittent vibration image), while for N=3, there is a very small positive spike in the energy density at $z = l$, so only unclear and unstable standing wave images appear near the axis of symmetry. However, for *N*=2 and N=4, there are large positive spikes at $z = l$, and the energy density distribution has standing waves or forms clear and stable intermittent vibration images.

From Figures 2-4 and Equation (46), it can be seen that the energy density of the fluid needs to be sampled and added at finite sampling intervals $\Delta z$, where $\Delta z = \bar{u}_z \Delta t \approx V_0 \Delta t$ and $\bar{u}_z \approx V_0$ is a constant. Through numerical computation we find that only when $\Delta t$ satisfies

$$\Delta t \approx \frac{9.54 \pm 0.01}{V_0} \tag{53}$$

can the stable and clear intermittent vibration patterns be obtained.

As is well known, the Dirac function is a generalized function, but it can be approximated (in the sense of weak convergence [9]) using a sequence of elementary functions, such as

$$\delta_\nu(z) = \frac{\sin \nu z}{\pi z} \tag{54}$$

and

$$\delta(z) = \delta_\nu(z)\big|_{\nu \to \infty} \tag{55}$$

Therefore, we roughly represent energy density as

$$E_s(r,z) = \frac{1}{\nu}[A_1 \delta_\nu(z) + A_2 \delta_\nu(l-z)]_{\nu \gg 1} = \frac{1}{\nu}[A_1 \frac{\sin \nu z}{\pi z} + A_2 \frac{\sin \nu(l-z)}{\pi(l-z)}]_{\nu \gg 1} \tag{56}$$

Each term in equation (56) can be regarded as a sinusoidal standing wave distribution with varying amplitudes. The former decreases with the increase of z, while the latter increases with the increase of z. Building upon the previous discussion, their standing wave images can be represented by several standing wave units and one 1/4 wavelength oscillator. Substituting (56) into (46) to get

$$E_m(r,z;K) = \sum_{k=0}^{K} \frac{1}{\nu}\{A_1 \delta_\nu(z+k\Delta z) + A_2 \delta_\nu(l-z-k\Delta z)\} \tag{57}$$

where *K* equals to the largest integer of $\dfrac{l}{\Delta z}$, i.e.

$$K = [\dfrac{l}{\Delta z}] \tag{58}$$

Obviously, equation (57) represents the sum of comb pulses or sampling pulses. If an appropriate value is selected to make the comb pulses added synchronously, their standing wave images should correspond to the case described in equations (50) - (52), that is to say, they should include several standing wave units and one 1/4 wavelength oscillator.

Now let's study the radial distribution of static energy density in the tube. Figures 5 and 6 represent the distribution of energy density along the radial direction in the tube when N=2 and N=4, respectively, where *x* is the radial coordinate variables. The results in the figures above-mentioned indicate that the static energy density is the smallest near the axis of symmetry. As the radial coordinates increase, the energy density increases until it reaches its maximum value in the neighborhood of $x = 0.85$ and then decreases. Figure 7 shows the distribution of energy density along the radius direction for N=3. It can be seen from this figure that, unlike the cases of N=2 and N=4, the static energy density reaches its maximum near the symmetry axis and then decreases, with a small positive spike at $z = l$ then disappearing near x=0.25.

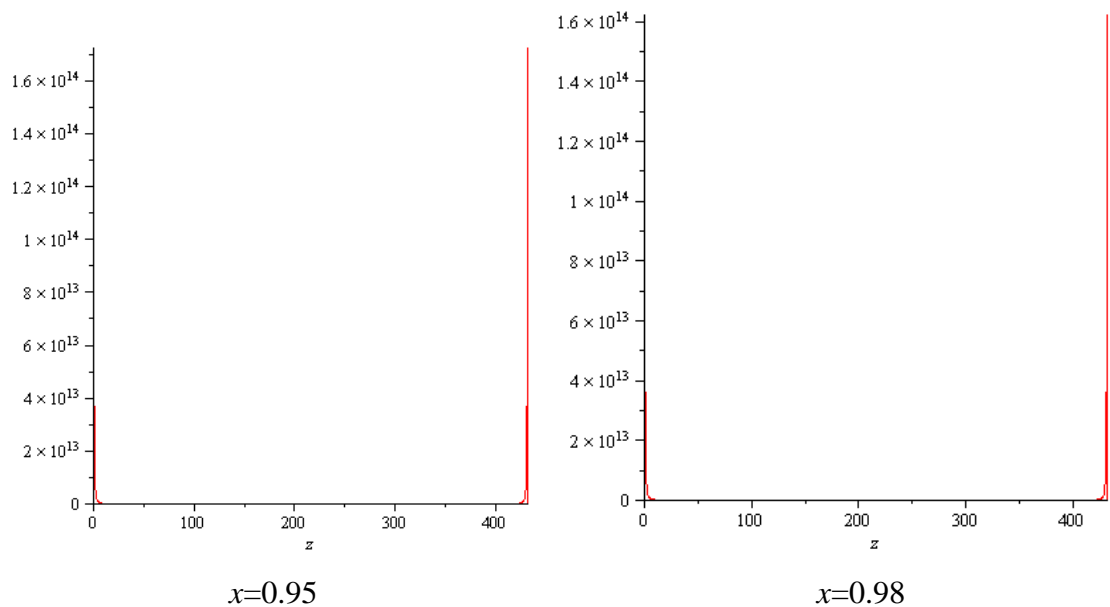

x=0.95　　　　　　　　　　　　　　x=0.98

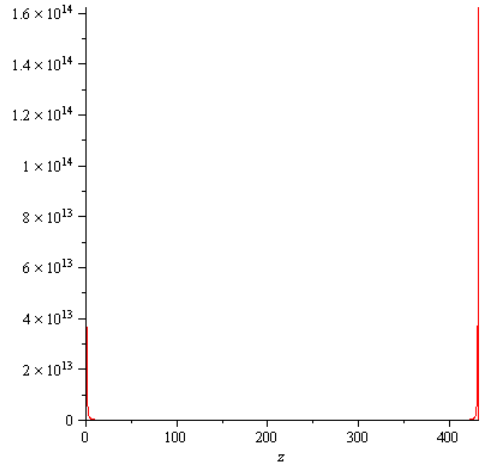

*x*=0.75

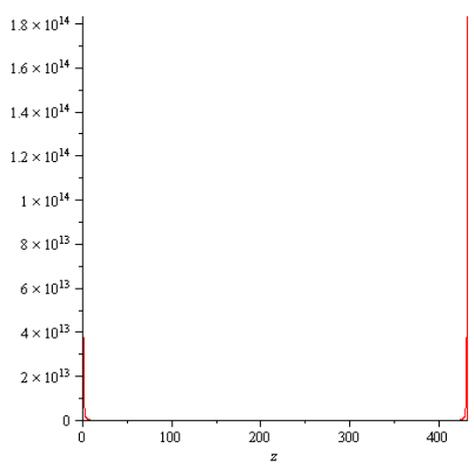

*x*=0.85

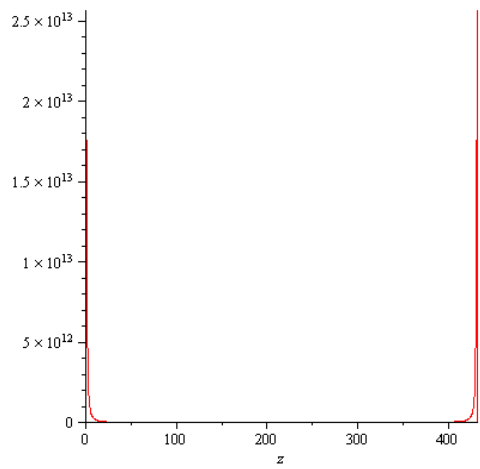

*x*=0.25

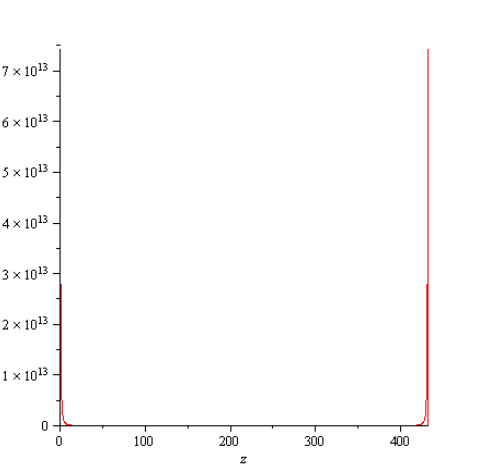

*x*=0.5

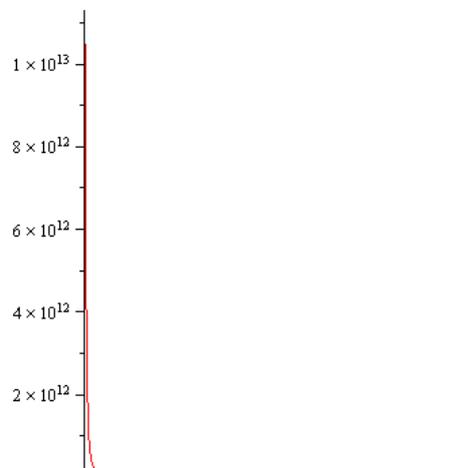

*x*=0

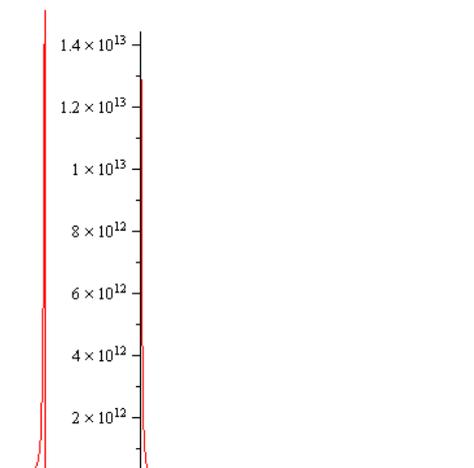

*x*=0.1

Figure 5. The static radial distribution of energy density along the for N=2, *x* is the radial coordinate variables.

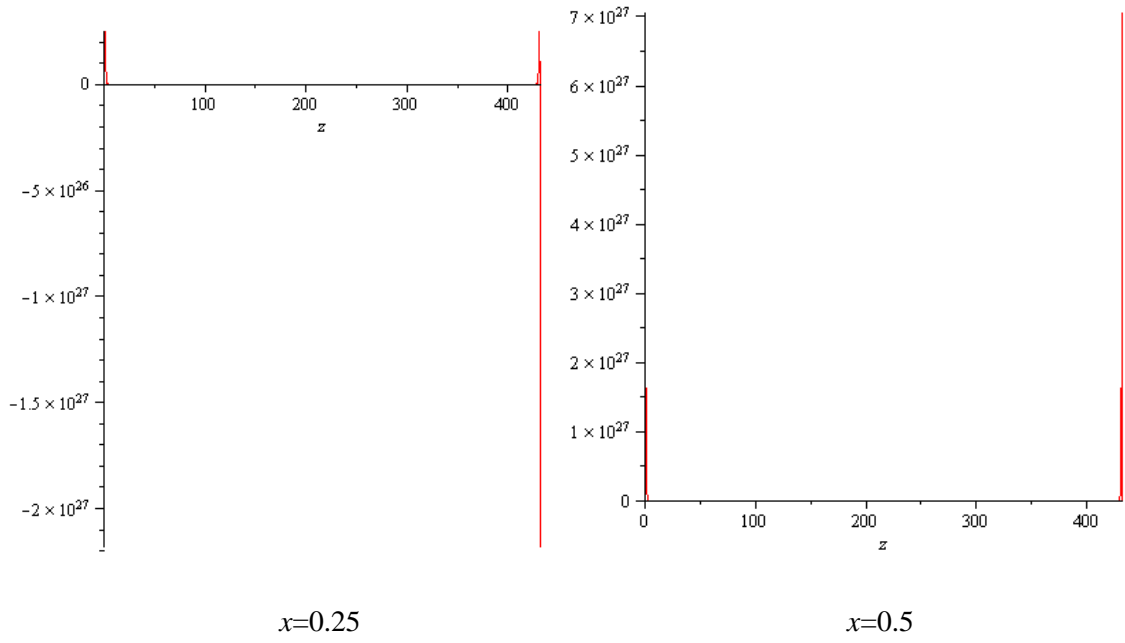

*x*=0.25            *x*=0.5

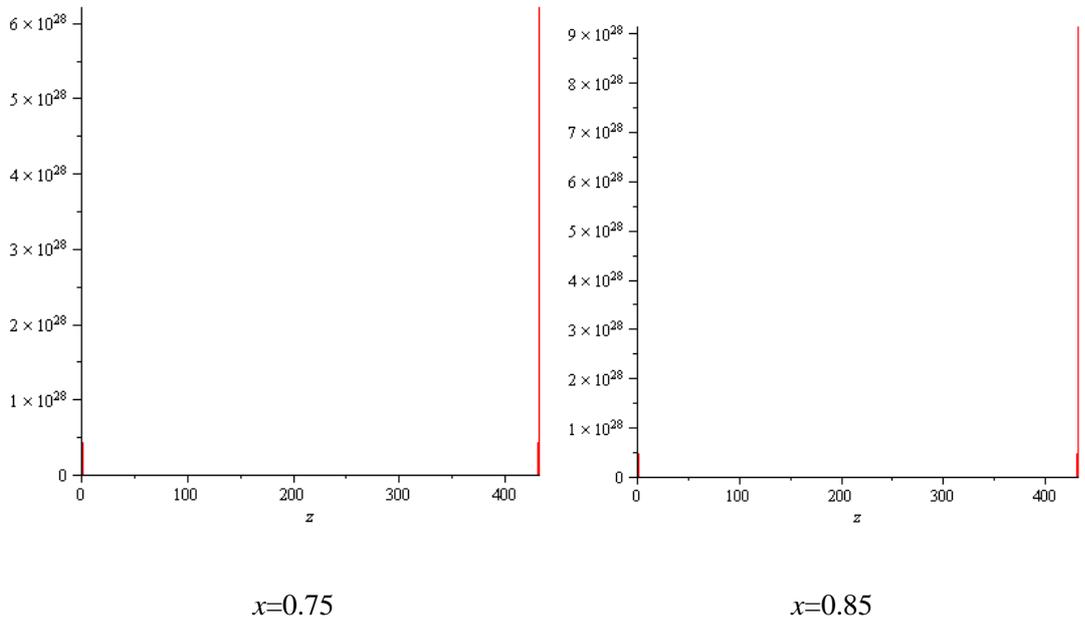

*x*=0.75            *x*=0.85

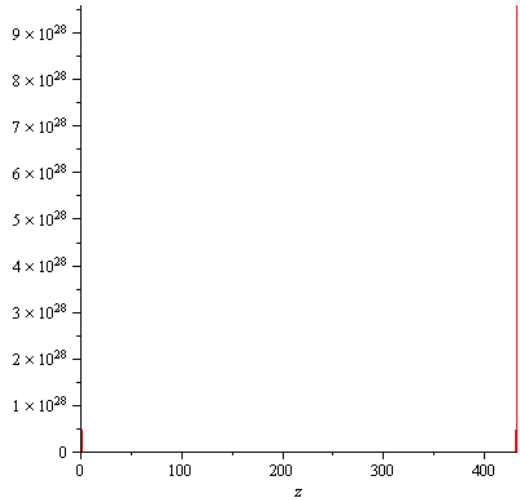

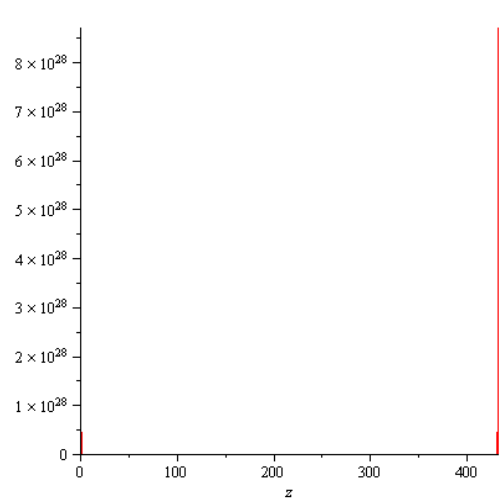

*x*=0.9

*x*=0.95

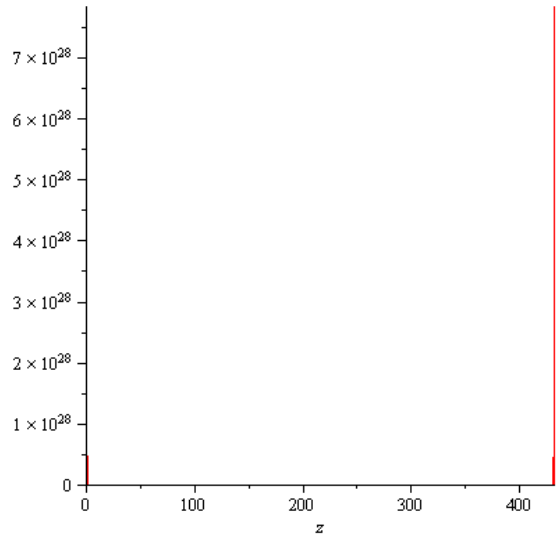

*x*=0.8

Figure 6. The static radial distribution of energy density along the for N=4, *x* is the radial coordinate variables.

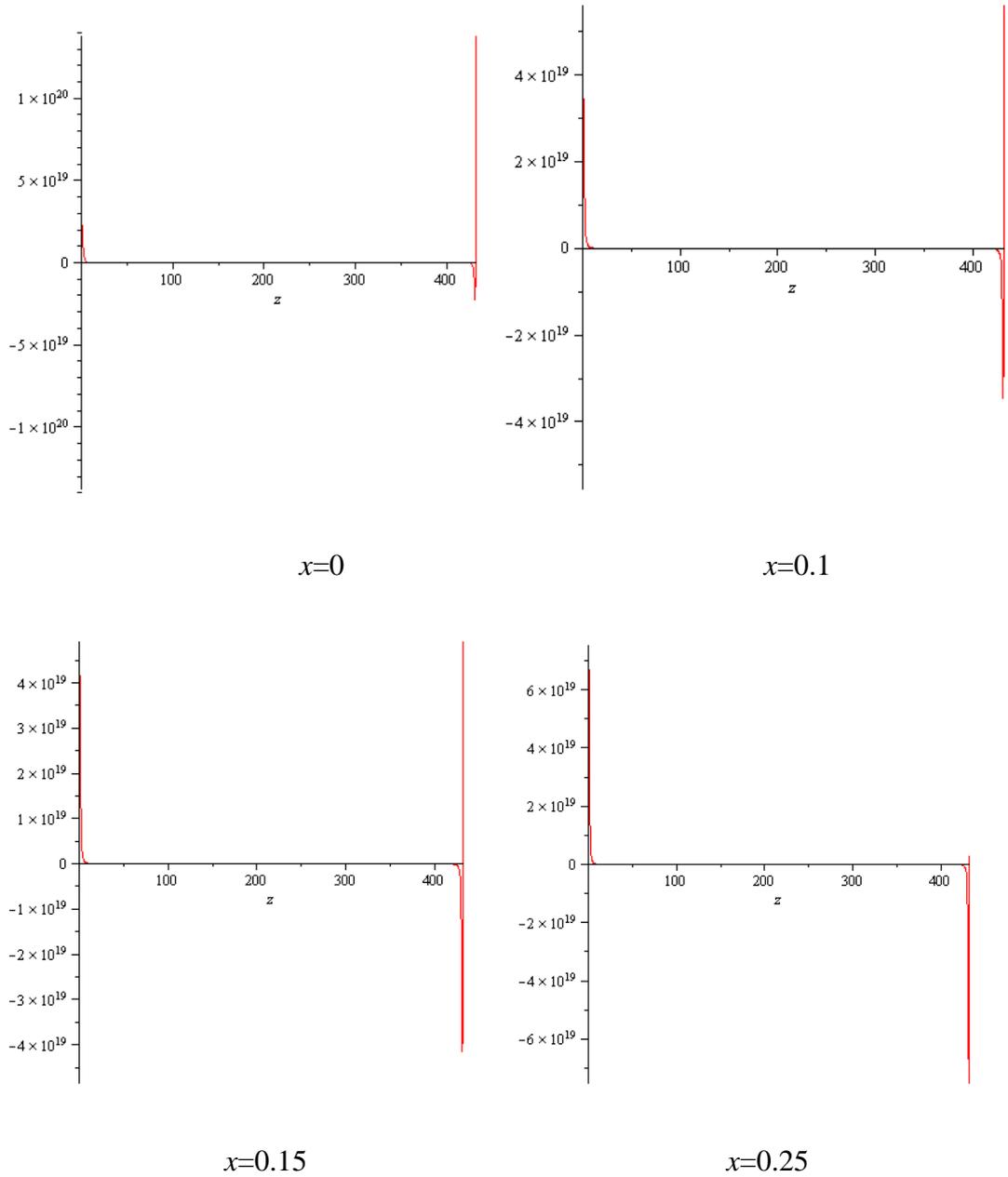

Figure 7. The static radial distribution of energy density along the for N=3, *x* is the radial coordinate variables.

Based on the above discussion, we will now analyze the numerical computation results. Obviously, Figures 2- 4 represent the case in Equation (50), where two standing wave units are added with one 1/4 wavelength oscillator. Within an order of magnitude error range, the energy density of flowing fluid does not cross the line of 60 radii, which is consistent with Reynolds' experimental results.

In order to elucidate the progression of intermittent vibration, we fixed the sampling interval $\Delta z = 9.53$ and $x=0.5$ in (46) for $N=4$ to execute numerical calculation, and the obtained result is shown in Figure 8. The first image $K=2$ shows one 1/4 wavelength oscillator; The second image, $K=20$, shows one standing wave unit and one 1/4 wavelength oscillator; The third figure $K=35$ shows two standing wave units plus one 1/4 wavelength oscillator; The fourth to sixth images show $K=45$, 50, and 55, respectively, showing stable final images with two standing wave units and one 1/4 wavelength oscillator. The above interval $\Delta z = 9.53$ indicates that it is a synchronous sampling interval.

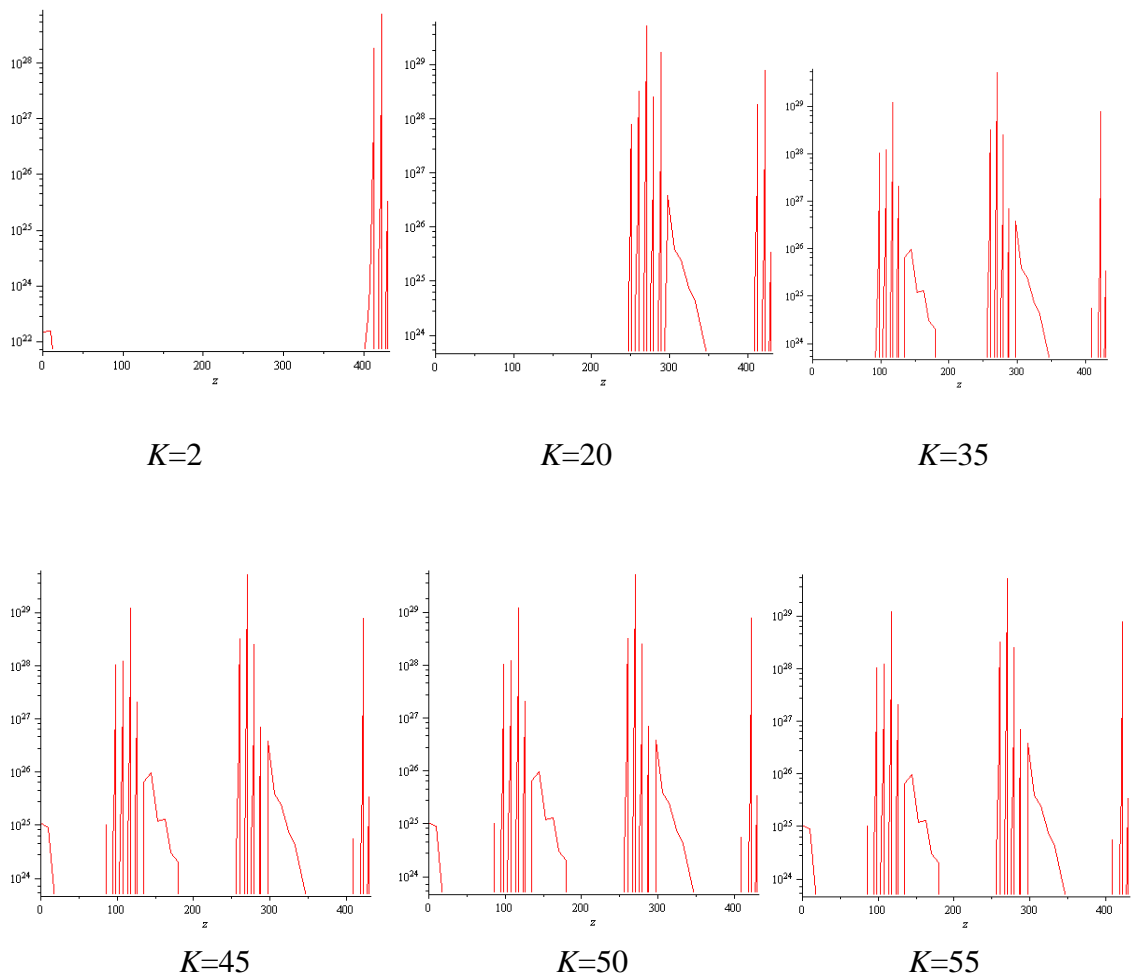

| $K=2$ | $K=20$ | $K=35$ |
| $K=45$ | $K=50$ | $K=55$ |

Figure 8. Progress process of intermittent vibration for $N=4$.

**Conclusion**

Through numerical computations, we found that only when the sampling time interval $\Delta t$ satisfies

$$\Delta t \approx \frac{9.54 \pm 0.01}{V_0}$$

then stable and clear intermittent vibration patterns can be obtained as shown in Figures 2-4, from which we can see that the intermittent vibration images obtained in this way are consistent with those observed by Reynolds in a cylindrical tube before 140 years.

On the other hand, it can be inferred from equations (44) and (A5.13) - (A5.17) that,

$$E_s(r,z) \propto \frac{1}{(\alpha R)^N} \tag{59}$$

with $\frac{1}{\alpha R} \propto 10^6 \text{-} 10^7$, which means that as $N$ increases it increases until it reaches infinity. For the first time in history, a complete explanation of Reynolds' experiments is provided.

## Appendix 1

Navier-Stokes' equation

$$\frac{\partial \boldsymbol{u}}{\partial t} + (\boldsymbol{u} \cdot \nabla)\boldsymbol{u} = -\frac{1}{\rho}\nabla P + \nu\nabla^2 \boldsymbol{u} \tag{A1.1}$$

Carrying out a curl operation on equation (A1.1) yields

$$\nabla \times \{\frac{\partial \boldsymbol{u}}{\partial t} + (\boldsymbol{u} \cdot \nabla)\boldsymbol{u}\} = \nabla \times \{-\frac{1}{\rho}\nabla P + \nu\nabla^2 \boldsymbol{u}\} \tag{A1.2}$$

Using of following vector formula

$$\nabla(a \cdot b) = a \times (\nabla \times b) + b \times (\nabla \times a) + (a \cdot \nabla)b + (b \cdot \nabla)a$$

one has

$$(\boldsymbol{u} \cdot \nabla)\boldsymbol{u} = -\boldsymbol{u} \times \nabla \times \boldsymbol{u} + \frac{1}{2}\nabla(u^2) \tag{A1.3}$$

Carrying out a curl operation on equation (A1.3) yields

$$\nabla \times [(\boldsymbol{u} \cdot \nabla)\boldsymbol{u}] = -\nabla \times [\boldsymbol{u} \times \nabla \times \boldsymbol{u}] \tag{A1.4}$$

Using of another vector formula

$$\nabla \times (a \times b) = a(\nabla \cdot b) - b(\nabla \cdot a) + (b \cdot \nabla)a - (a \cdot \nabla)b$$

We obtain

$$\nabla \times (\boldsymbol{u} \times \nabla \times \boldsymbol{u}) = (\nabla \times \boldsymbol{u} \cdot \nabla)\boldsymbol{u} - \nabla \times \boldsymbol{u}(\nabla \cdot \boldsymbol{u}) - (\boldsymbol{u} \cdot \nabla)\nabla \times \boldsymbol{u} + \boldsymbol{u}(\nabla \cdot \nabla \times \boldsymbol{u}).$$

For incompressible fluid, $\nabla \cdot \mathbf{u} = 0$ and the above formula becomes

$$\nabla \times (\mathbf{u} \times \nabla \times \mathbf{u}) = (\nabla \times \mathbf{u} \cdot \nabla)\mathbf{u} - (\mathbf{u} \cdot \nabla)\nabla \times \mathbf{u}$$

If the fluid is barotropic, substituting above expression into (A1.3), then put the obtained result in (A1.2), finally one has

$$\frac{\partial \nabla \times \mathbf{u}}{\partial t} - \nu\nabla^2(\nabla \times \mathbf{u}) = \{(\nabla \times \mathbf{u} \cdot \nabla)\mathbf{u} - (\mathbf{u} \cdot \nabla)(\nabla \times \mathbf{u})\} \quad (A1.5)$$

For axisymmetric flow $\mathbf{u}$ is independent of $\vartheta$ and $\nabla \times \mathbf{u} = \mathbf{i}_\vartheta (\nabla \times \mathbf{u})_\vartheta$, we get

$$(\nabla \times \mathbf{u} \cdot \nabla)\mathbf{u} = (\nabla \times \mathbf{u})_\vartheta \cdot \frac{\partial \mathbf{u}}{r\partial \vartheta} = 0$$

then (A1.5) becomes

$$\frac{\partial \nabla \times \mathbf{u}}{\partial t} - \nu\nabla^2(\nabla \times \mathbf{u}) = -(\mathbf{u} \cdot \nabla)(\nabla \times \mathbf{u}) \quad (A1.6)$$

On the other hand, carrying out a divergence operation on equation (A1.1) yields

$$\frac{\partial \nabla \cdot \mathbf{u}}{\partial t} + \nabla \cdot \{(\mathbf{u} \cdot \nabla)\mathbf{u}\} = -\nabla \cdot \{\frac{1}{\rho}\nabla P\} + \nabla \cdot \{\nu\nabla^2 \mathbf{u}\} \quad (A1.7)$$

and executing a divergence operation on equation (A1.3) yields

$$\nabla \cdot \{(\mathbf{u} \cdot \nabla)\mathbf{u}\} = \frac{1}{2}\nabla^2(u^2) - \nabla \cdot [\mathbf{u} \times \nabla \times \mathbf{u}] \quad (A1.8)$$

Using the vector formula

$$\nabla \cdot (a \times b) = b \cdot (\nabla \times a) - a \cdot (\nabla \times b)$$

we get

$$\nabla \cdot \{(\mathbf{u} \cdot \nabla)\mathbf{u} = \frac{1}{2}\nabla^2(u^2) - (\nabla \times \mathbf{u})^2 + \mathbf{u} \cdot \{\nabla\nabla \cdot \mathbf{u} - \nabla^2 \mathbf{u}\}$$

Substituting this result into (A1.7), one has

$$\frac{\partial \nabla \cdot \mathbf{u}}{\partial t} - \{\nu\nabla^2(\nabla \cdot \mathbf{u})\} = -\nabla^2(\frac{1}{2}V^2 + \frac{P}{\rho}) + (\nabla \times \mathbf{u})^2 - \mathbf{u} \cdot \{\nabla\nabla \cdot \mathbf{u} - \nabla^2 \mathbf{u}\} - \nabla \cdot \{P\nabla(\frac{1}{\rho})\}$$

For incompressible fluid, above expression can be approximately

$$\nabla^2(\frac{1}{2}V^2 + \frac{P}{\rho}) = (\nabla \times \mathbf{u})^2 + \mathbf{u} \cdot \nabla^2 \mathbf{u} \quad (A1.9)$$

$$B_2(r)=(\varepsilon A^{(0)})^2\{(\frac{\partial}{\partial r}+\frac{1}{r})[f^{(1)}(r)J_0(\alpha Rr)]\}\int f^{(1)}(r)J_0(\alpha Rr)dr$$

$$(\frac{\partial}{\partial r}+\frac{1}{r})[f^{(1)}(r)J_0(\alpha Rr)] \approx (\frac{\partial}{\partial r}+\frac{1}{r})[\frac{1}{4}r^2] = \frac{1}{4}(2r+r) = \frac{3}{4}r \qquad (A1.5)$$

$$\int f^{(1)}(r)J_0(\alpha Rr)dr = \int \frac{1}{4}r^2 dr = \frac{1}{12}r^3$$

$$B_2(r)=(\varepsilon A^{(0)})^2\{(\frac{\partial}{\partial r}+\frac{1}{r})[f^{(1)}(r)J_0(\alpha Rr)]\}\int f^{(1)}(r)J_0(\alpha Rr)dr$$
$$= (\varepsilon A^{(0)})^2 \frac{3}{4}r \cdot \frac{1}{12}r^3 = (\varepsilon A^{(0)})^2 \frac{1}{16}r^4 \qquad (A1.6)$$

Finally, one has

$$B(r) = B_1(r) + B_2(r) = \frac{1}{32}(\varepsilon A^{(0)})^2 r^4 + \frac{1}{16}(\varepsilon A^{(0)})^2 r^4 = \frac{3}{32}(\varepsilon A^{(0)})^2 r^4 \qquad (A1.7)$$

**Appendix 2. The solutions of equation (12)**

From Eqs. (10) and (12) in the text, one has

$$\nabla_r^2 \bar{\mathbf{\Omega}}^{(1)} + \kappa^2 \bar{\mathbf{\Omega}}^{(1)} = \varepsilon(1-r^2)\bar{\mathbf{\Omega}}^{(0)} = \mathbf{i}_g h^{(1)}(r)e^{-\alpha Rz}$$
$$h^{(1)}(r) = \varepsilon A^{(0)}(1-r^2)J_0(\kappa r) \qquad (A2.1)$$
$$\text{Re}\,\alpha R = \varepsilon$$

and the homogeneous solution of (A2.1) can be written as

$$\bar{\mathbf{\Omega}}^{(0)} = A^{(0)} J_0(\kappa Rr)e^{-\alpha Rz}$$
$$\kappa = \sqrt{\alpha^2 - \text{Re}\,\gamma} \qquad (A2.2)$$

where $A^{(0)}$ is a constant. Now we use the parameter variation method to solve the first equation (A2.1), let

$$\bar{\mathbf{\Omega}}^{(1)} = \mathbf{i}_g A^{(1)}(r) J_0(\kappa r)e^{-\alpha Rz} \qquad (A2.3)$$

Substituting (A2.3) into (A2.1) yields

$$\{\frac{d^2 A^{(1)}(r)}{dr^2}J_0(\kappa r) + \frac{dA^{(1)}(r)}{dr}[2\frac{d}{dr}J_0(\kappa r) + \frac{1}{r}J_0(\kappa r)]\} = h^{(1)}(r) \qquad (A2.4)$$

In the calculation, we applied the result that the $J_0(\kappa r)e^{-\alpha z}$ satisfies to the homogeneous equation of (A2.2), that is

$$\frac{d^2}{dr^2}J_0(\kappa r)+\frac{1}{r}\frac{d}{dr}J_0(\kappa r)+\kappa^2 J_0(\kappa r)=0$$

Now let's find the special solution of equation (A2.4). Let

$$\frac{d}{dr}A^{(1)}(r)=f \qquad (A2.5)$$

By substitution of (A2.5) into (A2.4), one has

$$f'(r)J_0(\kappa r)+f(r)[2\frac{d}{dr}J_0(\kappa r)+\frac{1}{r}J_0(\kappa r)]=h^{(1)}(r) \qquad (A2.6)$$

Let the right end of the above equation be equal to zero, we obtain the corresponding homogeneous equation whose solution is

$$f(r)=f_0\exp\{-\int[-2\kappa\frac{J_1(\kappa r)}{J_0(\kappa r)}+\frac{1}{r}]dr\}=f_0\frac{1}{r[J_0(\kappa r)]^2} \qquad (A2.7)$$

where $f_0$ is a constant. Using the method of parameter variation to find the non-homogeneous solution of equation (A2.6), which can be written as

$$f(r)=f_0(r)\frac{1}{r[J_0(\kappa r)]^2} \qquad (A2.8)$$

Substituting (A2.8) into (A2.6) yields

$$f'(r)J_0(\kappa r)+f(r)[-2\kappa J_1(\kappa r)+\frac{1}{r}J_0(\kappa r)]=f_0'(r)\frac{1}{rJ_0(\kappa r)} \qquad (A2.9)$$

By comparing (A2.9) with (A2.6), it is easy to obtain

$$f_0'(r)=rh^{(1)}(r)J_0(\kappa r)$$

or

$$f_0(r)=\int h^{(1)}(r)J_0(\kappa r)rdr \qquad (A2.10)$$

$$f(r)=\frac{dA^{(1)}(r)}{dr}=\frac{1}{r[J_0(\kappa r)]^2}\int h^{(1)}(r)J_0(\kappa r)rdr$$

$$A^{(1)}(r)=\int\frac{dr}{r[J_0(\kappa r)]^2}\int h^{(1)}(r)J_0(\kappa r)rdr$$
$$=\varepsilon A^{(0)}\int\frac{dr}{r[J_0(\kappa r)]^2}\int(1-r^2)J_0(\kappa r)J_0(\kappa r)rdr \qquad (A2.11)$$

$$\Omega^{(1s)}(r,z)=A^{(1)}(r)J_0(\alpha Rr)e^{-\alpha Rz}=\varepsilon A^{(0)}f^{(1)}(r)J_0(\alpha Rr)e^{-\alpha Rz} \qquad (A2.12)$$

$$f^{(1)}(r) = \int \frac{dr}{r[J_0(\alpha Rr)]^2} \int (1-r^2) J_0^2(\alpha Rr) r dr \tag{A2.13}$$

By means of Schafheitlin's reduction formula[10]

$$(\mu+2)\int^z z^{\mu+2} J_\nu^2(z) dz = (\mu+1)[\nu^2 - \frac{1}{4}(\mu+1)^2] \int^z z^\mu J_\nu^2(z) dz + \frac{1}{2}[z^{\mu+1}\{z J_\nu'(z) - \frac{1}{2}(\mu+1) J_\nu(z)\}^2 + z^{\mu+1}\{z^2 - \nu^2 + \frac{1}{4}(\mu+1)^2\} J_\nu^2(z)] \tag{A2.14}$$

When $\mu = 2m+1$, $\nu = 0$, and let

$$G(m,z) = \int z^{2m+3} [J_0(z)]^2 dz \tag{A2.15}$$

Then (A2.14) can be rewritten as

$$G(m,z) = -\frac{2(m+1)^3}{2m+3} G(m-1,z) + \frac{1}{2(2m+3)} z^{2m+2} \{[z J_0'(z) - \frac{1}{2}(2m+2) J_0(z)]^2 + [z^2 + \frac{1}{4}(2m+2)^2] J_0^2(z)\} \tag{A2.16}$$

Substituting equations (A2.14)- (A2.16) into (A2.13) yields

$$f^{(1)}(r) = \int \frac{dr}{r[J_0(\alpha r)]^2} \int \frac{r}{R^2} (1-\frac{r^2}{R^2}) J_0^2(\alpha r) dr = \int \frac{dx}{x[J_0(x)]^2} \int \frac{x}{X^2} (1-\frac{x^2}{X^2}) J_0^2(x) dx$$

$$= \frac{1}{X^2} \int \frac{dx}{x[J_0(x)]^2} G(-1,x) - \frac{1}{X^4} \int \frac{dx}{x[J_0(x)]^2} G(0,x) \tag{A2.17}$$

where

$$X = \alpha R, \ x = \alpha r. \tag{A2.18}$$

From (A2.15), one has

$$G(-1,x) = \frac{1}{2} x^2 J_0^2(x) [1 + \frac{J_1^2(x)}{J_0^2(x)}]$$

$$G(0,x) = \frac{1}{6} x^2 J_0^2(x) \{(x^2 - 2)[\frac{J_1^2(x)}{J_0^2(x)} + 1] + 2[1 + x \frac{J_1(x)}{J_0(x)}]\} \tag{A2.19}$$

and

$$f^{(1)}(r) = \frac{1}{X^2} \int \frac{dx}{x[J_0(x)]^2} G(-1,x) - \frac{1}{X^4} \int \frac{dx}{x[J_0(x)]^2} G(0,x)$$

$$= \frac{1}{X^2} \int \frac{\frac{1}{2}x^2 J_0^2(x)[1 + \frac{J_1^2(x)}{J_0^2(x)}]dx}{x[J_0(x)]^2}$$

$$- \frac{1}{X^4} \int \frac{\frac{1}{6}x^2 J_0^2(x)\{(x^2-2)[\frac{J_1^2(x)}{J_0^2(x)} + 1] + 2[1 + x\frac{J_1(x)}{J_0(x)}]\}dx}{x[J_0(x)]^2} \qquad (A2.20)$$

$$= \frac{1}{2X^2} \int x[1 + \frac{J_1^2(x)}{J_0^2(x)}]dx - \frac{1}{6X^4} \int x\{(x^2-2)[\frac{J_1^2(x)}{J_0^2(x)} + 1] + 2[1 + x\frac{J_1(x)}{J_0(x)}]\}dx$$

where

$$J_0(x) = 1 - \frac{1}{4}x^2 + \frac{1}{64}x^4, \quad J_1(x) = \frac{1}{2}x - \frac{1}{16}x^3$$

$$\frac{J_1(x)}{J_0(x)} = \frac{\frac{1}{2}x(1-\frac{1}{8})x^3}{1 - \frac{1}{4}x^2 + \frac{1}{64}x^4} \approx \frac{1}{2}x(1 - \frac{1}{8}x^2)(1 + \frac{1}{4}x^2 - \frac{1}{64}x^4 + \frac{1}{16}x^4) \approx \frac{1}{2}x(1 + \frac{1}{4}x^2 + \frac{3}{64}x^4 - \frac{1}{8}x^2 - \frac{1}{32}x^4)$$

$$= \frac{1}{2}x(1 + \frac{1}{8}x^2 + \frac{1}{64}x^4)$$

$$1 + x\frac{J_1(x)}{J_0(x)} \approx 1 + \frac{1}{2}x^2(1 + \frac{1}{8}x^2 + \frac{1}{64}x^4)$$

$$[\frac{J_1(x)}{J_0(x)}]^2 = \frac{[\frac{1}{2}x(1 + \frac{1}{8}x^2 + \frac{1}{64}x^4)]^2}{[1 - \frac{1}{4}x^2 + \frac{1}{64}x^4]^2} \approx \frac{1}{4}x^2(1 + \frac{1}{4}x^2 + \frac{1}{32}x^4 + \frac{1}{64}x^4)[1 + 2(\frac{1}{4}x^2 - \frac{1}{64}x^4) + 3(\frac{1}{4}x^2 - \frac{1}{64}x^4)^2]$$

$$\approx \frac{1}{4}x^2(1 + \frac{1}{4}x^2 + \frac{3}{64}x^4)[1 + 2(\frac{1}{4}x^2 - \frac{1}{64}x^4) + \frac{3}{16}x^4]$$

$$\approx \frac{1}{4}x^2(1 + \frac{1}{4}x^2 + \frac{3}{64}x^4)[1 + (\frac{1}{2}x^2 + \frac{5}{32}x^4)]$$

$$\approx \frac{1}{4}x^2[1 + (\frac{1}{2}x^2 + \frac{5}{32}x^4) + \frac{1}{4}x^2(1 + \frac{1}{4}x^2)] = \frac{1}{4}x^2[1 + (\frac{3}{4}x^2 + \frac{7}{32}x^4)]$$

$$1 + [\frac{J_1(x)}{J_0(x)}]^2 = 1 + \frac{1}{4}x^2[1 + (\frac{3}{4}x^2 + \frac{7}{32}x^4)]$$

Substituting these results into (A2.19) yields

$$G(-1,x) = \int x[J_0(x)]^2 dx = \frac{1}{2}x^2[J_1^2(x)+J_0^2(x)] = \frac{1}{2}x^2 J_0^2(x)[1+\frac{J_1^2(x)}{J_0^2(x)}]$$

$$\approx \frac{1}{2}x^2 J_0^2(x)\{1+\frac{1}{4}x^2[1+(\frac{3}{4}x^2+\frac{7}{32}x^4)]\}$$

$$G(0,x) = \int x^3[J_0(x)]^2 dx = \frac{1}{6}x^2(x^2-2)[J_1^2(x)+J_0^2(x)] + \frac{1}{3}x^2[J_0^2(x)+xJ_1(x)J_0(x)]$$

$$= \frac{1}{6}x^2 J_0^2(x)\{(x^2-2)[\frac{J_1^2(x)}{J_0^2(x)}+1]+2[1+x\frac{J_1(x)}{J_0(x)}]\}$$

$$\approx \frac{1}{6}x^2 J_0^2(x)(x^2-2)\{1+\frac{1}{4}x^2[1+(\frac{3}{4}x^2+\frac{7}{32}x^4)]\}+2\cdot\frac{1}{6}x^2 J_0^2(x)\{1+\frac{1}{2}x^2(1+\frac{1}{8}x^2+\frac{1}{64}x^4)\}$$

and finally

$$f^{(1)}(r) = \int \frac{dr}{r[J_0(\alpha r)]^2}\int \frac{r}{R^2}(1-\frac{r^2}{R^2})J_0^2(\alpha r)dr = \int \frac{dx}{x[J_0(x)]^2}\int \frac{x}{X^2}(1-\frac{x^2}{X^2})J_0^2(x)dx$$

$$= \frac{1}{X^2}\int \frac{\frac{1}{2}x^2 J_0^2(x)\{1+\frac{1}{4}x^2[1+(\frac{3}{4}x^2+\frac{7}{32}x^4)]\}dx}{x[J_0(x)]^2}$$

$$-\frac{1}{X^4}\int \frac{\frac{1}{6}x^2 J_0^2(x)(x^2-2)\{1+\frac{1}{4}x^2[1+(\frac{3}{4}x^2+\frac{7}{32}x^4)]\}dx+2\cdot\frac{1}{6}x^2 J_0^2(x)\{1+\frac{1}{2}x^2(1+\frac{1}{8}x^2+\frac{1}{64}x^4)\}dx}{x[J_0(x)]^2}$$

$$= \frac{1}{X^2}\int\{\frac{1}{2}x+\frac{1}{8}x^3+\frac{3}{32}x^5+\frac{7}{256}x^7\}dx$$

$$-\frac{1}{X^4}\int\frac{1}{6}x(x^2-2)\{1+\frac{1}{4}x^2[1+(\frac{3}{4}x^2+\frac{7}{32}x^4)]\}dx - \frac{1}{X^4}\int 2\cdot\frac{1}{6}x\{1+\frac{1}{2}x^2(1+\frac{1}{8}x^2+\frac{1}{64}x^4)\}dx$$

$$= \frac{1}{X^2}\int\{\frac{1}{2}x+\frac{1}{8}x^3+\frac{3}{32}x^5+\frac{7}{256}x^7\}dx$$

$$-\frac{1}{X^4}\int\{[\frac{1}{6}x^3(1+\frac{1}{4}x^2+\frac{3}{16}x^4+\frac{7}{128}x^6)-2\cdot\frac{1}{6}x(1+\frac{1}{4}x^2+\frac{3}{16}x^4+\frac{7}{128}x^6)]+(\frac{1}{3}x+\frac{1}{6}x^3+\frac{1}{48}x^5+\frac{1}{384}x^7)$$

$$= \frac{1}{X^2}\int\{\frac{1}{2}x+\frac{1}{8}x^3+\frac{3}{32}x^5+\frac{7}{256}x^7\}dx - \frac{1}{X^4}\int\{\frac{1}{12}x^3-\frac{1}{24}x^5+\frac{1}{64}x^7\}dx$$

$$\approx \frac{1}{4}r^2(1-\frac{1}{24}r^2)+O(\alpha^2 R^2)$$

(A2.21)

## Appendix 3 Velocity field

Since the motion is axisymmetric, one has

$$\nabla \times \boldsymbol{u}^{(1)} = \frac{1}{r}\begin{vmatrix} \mathbf{i}_r & r\mathbf{i}_\vartheta & \mathbf{i}_z \\ \frac{\partial}{\partial r} & \frac{\partial}{\partial \vartheta} & \frac{\partial}{\partial z} \\ u_r^{(1)} & 0 & u_z^{(1)} \end{vmatrix} = \mathbf{i}_\vartheta(\frac{\partial u_r^{(1)}}{\partial z} - \frac{\partial u_z^{(1)}}{\partial r}) = \mathbf{i}_\vartheta A^{(1)}(r)J_0(\kappa Rr)e^{-\alpha Rz} \quad (A3.1)$$

Differentiating (A3.1) with respect to $z$ and $r$, respectively, we get

$$\frac{\partial^2 u_r^{(1)}}{\partial z^2} - \frac{\partial^2 u_z^{(1)}}{\partial z \partial r} = -\alpha A^{(1)}(r)J_0(\kappa Rr)e^{-\alpha Rz} \quad (A3.2)$$

and

$$\frac{\partial^2 u_r^{(1)}}{\partial r \partial z} - \frac{\partial^2 u_z^{(1)}}{\partial r^2} = \frac{\partial}{\partial r}[A^{(1)}(r)J_0(\kappa Rr)]e^{-\alpha Rz} \quad (A3.3)$$

On the other hand, since the fluid is incompressible, one has

$$\frac{1}{r}\frac{\partial}{\partial r}\left(ru_r^{(1)}\right) + \frac{\partial u_z^{(1)}}{\partial z} = 0 \quad (A3.4)$$

Differentiating (A3.4) with respect to $r$, we get

$$\frac{\partial}{\partial r}[\frac{1}{r}\frac{\partial}{\partial r}\left(ru_r^{(1)}\right)] + \frac{\partial^2 u_z^{(1)}}{\partial r \partial z} = 0 \quad (A3.5)$$

Adding (A3.2) and (A3.5) yields

$$\frac{\partial}{\partial r}[\frac{1}{r}\frac{\partial}{\partial r}\left(ru_r^{(1)}\right)] + \frac{\partial^2 u_r^{(1)}}{\partial z^2} = [\frac{\partial^2}{\partial r^2} + \frac{\partial}{r\partial r} - \frac{1}{r^2} + \frac{\partial^2}{\partial z^2}]u_r^{(1)} = -\alpha A^{(1)}(r)J_0(\kappa Rr)e^{-\alpha Rz} \quad (A3.6)$$

On the other hand, since

$$\nabla \times \nabla \times \boldsymbol{u}^{(1)} = \frac{1}{r}\begin{vmatrix} \mathbf{i}_r & r\mathbf{i}_\vartheta & \mathbf{i}_z \\ \frac{\partial}{\partial r} & \frac{\partial}{\partial \vartheta} & \frac{\partial}{\partial z} \\ 0 & rA^{(1)}(r)J_0(\kappa r)e^{-\alpha Rz} & 0 \end{vmatrix} = \\ = -\mathbf{i}_r\frac{\partial}{\partial z}\{A^{(1)}(r)J_0(\kappa Rr)e^{-\alpha Rz}\} + \mathbf{i}_z\frac{1}{r}\frac{\partial}{\partial r}\{rA^{(1)}(r)J_0(\kappa Rr)\}e^{-\alpha Rz} \quad (A3.7)$$

and the fluid is an incompressible, one has

$$\nabla \times \nabla \times \boldsymbol{u} = \nabla\nabla \cdot \boldsymbol{u} - \nabla^2 \boldsymbol{u} = -\nabla^2 \boldsymbol{u}$$

then

$$\nabla^2 \boldsymbol{u}^{(1)} = -\{\mathbf{i}_r 2\alpha A^{(1)}(r)J_0(\kappa Rr) + \mathbf{i}_z(\frac{\partial}{\partial r} + \frac{1}{r})A^{(1)}(r)J_0(\kappa Rr)\}e^{-\alpha Rz} = \mathbf{i}_r\nabla^2 u_r^{(1)} + \mathbf{i}_z\nabla^2 u_z^{(1)}$$

Then we get

$$\nabla^2 u_r^{(1)} = -\alpha A^{(1)}(r) J_0(\kappa Rr) e^{-\alpha Rz}$$

$$\nabla^2 u_z^{(1)} = -(\frac{\partial}{\partial r} + \frac{1}{r}) A^{(1)}(r) J_0(\kappa r) e^{-\alpha z}$$

or

$$[(\frac{\partial^2}{\partial r^2} + \frac{\partial}{r\partial r}) + \frac{\partial^2}{\partial z^2}] u_r^{(1)} = -2\alpha A^{(1)}(r) J_0(\kappa r) e^{-2\alpha z} \tag{A3.8}$$

$$[(\frac{\partial^2}{\partial r^2} + \frac{\partial}{r\partial r}) + \frac{\partial^2}{\partial z^2}] u_z^{(1)} = -(\frac{\partial}{\partial r} + \frac{1}{r}) A^{(1)}(r) J_0(\kappa Rr) e^{-\alpha Rz} \tag{A3.9}$$

(A3.8) minus (A3.6), one has

$$\frac{1}{r^2} u_r^{(1)} = 0,$$

and

$$u_r^{(1)} = 0 \tag{A3.10}$$

Putting it in (A3.1) we get

$$-\frac{\partial u_z^{(1)}}{\partial r} = A^{(1)}(r) J_0(\kappa Rr) e^{-\alpha Rz}$$

By integration, one has

$$u_z^{(1)}(r,z) - u_z^{(1)}(0,z) = u_z^{(1)}(r,z) - u_z^{(1)}(0,z) = -e^{-\alpha z} \int_0^r A^{(1)}(r) J_0(\kappa Rr) dr$$

$$\nabla^2 u_z^{(1)} = -(\frac{\partial}{\partial r} + \frac{1}{r}) A^{(1)}(r) J_0(\kappa Rr) e^{-\alpha Rz} \tag{A3.11}$$

$$u_r^{(1)} = 0$$

$$\nabla^2 u_r^{(1)} = -\alpha A^{(1)}(r) J_0(\kappa Rr) e^{-\alpha Rz}$$

due to

$$u_z^{(1)}(0,z) = (1-r^2)_{r=0} = 1$$

one has

$$u_z^{(1)}(r,z) = u_z^{(1)}(0,z) - \int_0^r A^{(1)}(r) J_0(\kappa Rr) e^{-\alpha Rz} dr \approx 1 - \int_0^r A^{(1)}(r) J_0(\kappa Rr) dr \tag{A3.12}$$

Aat the wall of the tube $u_z^{(1)}(1,z) = 0$, i.e.

$$0 = 1 - \int_0^1 A^{(1)}(r) J_0(\kappa R r) dr = 1 - \varepsilon A^{(0)} \int_0^1 f^{(1)}(r) J_0(\kappa R r) dr$$

or

$$\varepsilon A^{(0)} = \frac{1}{\int_0^1 f^{(1)}(r) J_0(\kappa R r) dr} \quad (A3.13)$$

# Appendix 4

We rewrite the indefinite integral in equation (9) in reference [2] to the definite integral to obtain

$$u_z(r,z) - u_z(0,z) = -\int_0^r \Omega \, dr \quad (A4.1)$$

Substitute the equation (3) in reference [2] into (A4.1) and use the equation

$$u_z(0,z) = (1 - \frac{r^2}{R^2})\Big|_{r=0} e^{-\alpha R z} = e^{-\alpha R z} \quad (A4.2)$$

to obtain

$$u_z(r,z) = 1 - e^{-\alpha R z} \varepsilon A^{(0)} \int_0^r f^{(1)}(r) J_0(\alpha R r) dr \quad (A4.3)$$

According to fourth formula in equation (9) of reference [2]

$$\nabla^2 u_z = -(\frac{\partial}{\partial r} + \frac{1}{r})[A^{(1)}(r) J_0(\alpha R r)] e^{-\alpha R z} \quad (A4.4)$$

and

$$\boldsymbol{u} \cdot \nabla^2 \boldsymbol{u} = 0 + u_1 \cdot \nabla^2 u_z \quad (A4.5)$$

Substituting equations (A2.3) and (A2.4) into (A2.5) yields

$$\boldsymbol{u} \cdot \nabla^2 \boldsymbol{u} = B_2(r) e^{-2\alpha R z} \quad (A4.6)$$

Where(cf. equation (25) in the text)

$$B_2(r) = \{1 - \int \varepsilon A^{(0)} f^{(1)}(r) J_0(\alpha R r) dr\} \{-(\frac{\partial}{\partial r} + \frac{1}{r})[A^{(1)}(r) J_0(\alpha R r)]\} \quad (A4.7)$$

and (cf. equation (23) in the text)

$$B_1(r_s) = \frac{1}{2}[\varepsilon A^{(0)} f^{(1)}(r_s)]^2 J_0^2(\alpha R r_s) \tag{A4.8}$$

Oof which the calculation in detail as follows

$$\begin{aligned}
B_1(r_s) &= \frac{1}{2}[\varepsilon A^{(0)} f^{(1)}(r_s)]^2 J_0^2(\alpha R r_s) \approx \frac{1}{2}[\varepsilon A^{(0)} \frac{1}{4} r^2 (1 - \frac{1}{24} r^2)]^2 J_0^2(\alpha R r_s) \\
&\approx \frac{1}{2}(\varepsilon A^{(0)})^2 \frac{1}{16} r^4 (1 - \frac{1}{12} r^2 + \frac{1}{576} r^4)(1 - \frac{1}{2} r^2 + \frac{1}{16} r^4) \\
&= \frac{1}{32}(\varepsilon A^{(0)})^2 r^4 \{1 - \frac{1}{2} r^2 + \frac{1}{16} r^4 - \frac{1}{12} r^2 (1 - \frac{1}{2} r^2 + \frac{1}{16} r^4) \\
&\quad + \frac{1}{576} r^4 (1 - \frac{1}{2} r^2 + \frac{1}{16} r^4)\} \\
&= \frac{1}{32}(\varepsilon A^{(0)})^2 r^4 \{1 - \frac{7}{12} r^2 + \frac{61}{576} r^4 - \frac{7}{1152} r^6 + \frac{1}{9216} r^8\}
\end{aligned} \tag{A4.9}$$

and

$$B_2(r) = (\varepsilon A^{(0)})^2 \{(\frac{\partial}{\partial r} + \frac{1}{r})[f^{(1)}(r) J_0(\alpha R r)]\} \int f^{(1)}(r) J_0(\alpha R r) dr \tag{A4.10}$$

where

$$f^{(1)}(r) J_0(\alpha R r) = \frac{1}{4} r^2 (1 - \frac{1}{24} r^2)(1 - \frac{1}{4} r^2)$$

$$\begin{aligned}
f^{(1)}(r) J_0(\alpha R r) &= \frac{1}{4} r^2 (1 - \frac{1}{24} r^2)(1 - \frac{1}{4} r^2) = \frac{1}{4} r^2 (1 - \frac{1}{4} r^2 - \frac{1}{24} r^2 + \frac{1}{96} r^4) \\
&= \frac{1}{4} r^2 (1 - \frac{7}{24} r^2 + \frac{1}{96} r^4) = (\frac{1}{4} r^2 - \frac{7}{96} r^4 + \frac{1}{384} r^6)
\end{aligned}$$

$$\begin{aligned}
(\frac{\partial}{\partial r} + \frac{1}{r})[f^{(1)}(r) J_0(\alpha R r)] &= \frac{\partial}{\partial r}(\frac{1}{4} r^2 - \frac{7}{96} r^4 + \frac{1}{384} r^6) + \frac{1}{r}(\frac{1}{4} r^2 - \frac{7}{96} r^4 + \frac{1}{384} r^6) \\
&= \frac{1}{2} r - \frac{7}{24} r^3 + \frac{1}{64} r^5 + \frac{1}{4} r - \frac{7}{96} r^3 + \frac{1}{384} r^5 = \frac{3}{4} r - \frac{35}{96} r^3 + \frac{7}{384} r^5
\end{aligned}$$

$$\int f^{(1)}(r) J_0(\alpha R r) dr = \int (\frac{1}{4} r^2 - \frac{7}{96} r^4 + \frac{1}{384} r^6) dr = \frac{1}{12} r^3 - \frac{7}{480} r^5 + \frac{1}{7 \times 384} r^7$$

$$\varepsilon A^{(0)} = \frac{1}{\int_0^1 f^{(1)}(r) J_0(\alpha R r) dr} = \frac{1}{[\frac{1}{12} r^3 - \frac{7}{480} r^5 + \frac{1}{7 \times 384} r^7]_0^1} = \frac{13440}{929} \tag{A4.11}$$

$$B_2(r)=(\varepsilon A^{(0)})^2\{(\frac{\partial}{\partial r}+\frac{1}{r})[f^{(1)}(r)J_0(\alpha Rr)]\}\int f^{(1)}(r)J_0(\alpha Rr)dr$$

$$=(\varepsilon A^{(0)})^2(\frac{3}{4}r-\frac{35}{96}r^3+\frac{7}{384}r^5)(\frac{1}{12}r^3-\frac{7}{480}r^5+\frac{1}{7\times 384}r^7)$$

$$=(\varepsilon A^{(0)})^2[(\frac{1}{16}r^4-\{\frac{7}{640}+\frac{35}{96\times 12}\}r^6+\{\frac{1}{4\times 7\times 128}+\frac{245}{96\times 480}+\frac{7}{12\times 384}\}r^8 - \text{(A4.12)}$$

$$\{\frac{35}{96\times 7\times 384}+\frac{49}{480\times 384}\}r^{10}+\frac{7}{7\times 384\times 384}r^{12})]$$

$$=(\varepsilon A^{(0)})^2(\frac{1}{16}r^4-\frac{119}{2880}r^6+\frac{51}{7168}r^8-\frac{37}{92160}r^{10}+\frac{1}{147456}r^{12})$$

and

$$B(r)=B_1(r)+B_2(r)$$
$$=(\varepsilon A^{(0)})^2\{\frac{3}{32}r^4-\frac{343}{5760}r^6+\frac{1345}{129024}r^8-\frac{109}{184320}r^{10}+\frac{1}{98304}r^{12}\} \quad \text{(A4.13)}$$
$$=(\frac{929}{13440})^{-2}\{\frac{3}{32}r^4-\frac{343}{5760}r^6+\frac{1345}{129024}r^8-\frac{109}{184320}r^{10}+\frac{1}{98304}r^{12}\}$$

**Appendix 5**

In the text, we obtained the energy density integral

$$E(r,z)=e^{-2\alpha Rz}\int_0^1 B(r_s)r_s\{\int_0^{\psi_1}\frac{e^{2\alpha R|r_s-r|\tan\psi}}{\cos\psi}d\psi dr_s+\int_0^{\psi_2}\frac{e^{-2\alpha R|r_s-r|\tan\psi}}{\cos\psi}d\psi\}dr_s \quad \text{(A5.1)}$$

where

$$\tan\psi_1=\frac{z}{|r_s-r|}, \qquad \tan\psi_2=\frac{l-z}{|r_s-r|} \quad \text{(A5.2)}$$

Perform partial integration on the first expression in the right hand of the curly braces of equation (A1.1) to obtain

$$\int_0^{\psi_1} \frac{e^{2\alpha R|r_s-r|\tan\psi}}{\cos\psi} d\psi = \int_0^{\psi_1} \frac{\cos\psi e^{2\alpha R|r_s-r|\tan\psi}}{\cos^2\psi} d\psi = \int_0^{\psi_1} \cos\psi e^{2\alpha R|r_s-r|\tan\psi} d\tan\psi = \frac{1}{2\alpha R|r_s-r|} \int_0^{\psi_1} \cos\psi d e^{2\alpha R|r_s-r|\tan\psi}$$

$$= \frac{1}{2\alpha R|r_s-r|} \{\cos\psi e^{2\alpha R|r_s-r|\tan\psi} \Big|_0^{\psi_1} + \int_0^{\psi_1} \sin\psi e^{2\alpha R|r_s-r|\tan\psi} d\psi\}$$

$$= \frac{1}{2\alpha R|r_s-r|} \{\cos\psi_1 e^{2\alpha Rz} - 1 + \frac{1}{2\alpha R|r_s-r|} \int_0^{\psi_1} \cos^2\psi \sin\psi d e^{2\alpha R|r_s-r|\tan\psi}\}$$

$$= \frac{1}{2\alpha R|r_s-r|} [\frac{|r_s-r|}{\sqrt{|r_s-r|^2+z^2}} e^{2\alpha Rz} - 1] + \frac{1}{(2\alpha R|r_s-r|)^2} \int_0^{\psi_1} \cos^2\psi \sin\psi d e^{2\alpha R|r_s-r|\tan\psi}$$

Similarly, by performing partial integration on the second expression in the right hand of the curly braces of equation (A1.1) to get

$$\int_0^{\psi_2} \frac{e^{-2\alpha R|r_s-r|\tan\psi}}{\cos\psi} d\psi = \frac{-1}{2\alpha R|r_s-r|} [\frac{|r_s-r|}{\sqrt{|r_s-r|^2+(l-z)^2}} e^{-2\alpha R(l-z)} - 1] + [\frac{-1}{2\alpha R|r_s-r|}]^2 \int_0^{\psi_2} \cos^2\psi \sin\psi d e^{-2\alpha R|r_s-r|\tan\psi}$$

Substituting both results abovementioned into (A1.1) and using following approximation $2\alpha Rz \ll 1$, $e^{\pm 2\alpha Rz} \approx 1$

yield

$$\int_0^{\psi_1} \frac{e^{2\alpha R|r_s-r|\tan\psi}}{\cos\psi} d\psi + \int_0^{\psi_2} \frac{e^{-2\alpha R|r_s-r|\tan\psi}}{\cos\psi} d\psi_s \approx e^{-2\alpha Rz} \frac{1}{2\alpha R|r_s-r|} \{e^{2\alpha Rz}\cos\psi_1 - e^{-2\alpha R(l-z)}\cos\psi_2$$

$$+ \int_0^{\psi_1} \sin\psi e^{2\alpha R|r_s-r|\tan\psi} d\psi - \int_0^{\psi_2} \sin\psi e^{-2\alpha R|r_s-r|\tan\psi} d\psi\}$$

By same method continue to perform multiple partial integrals on the above equation finally obtain

$$\int_0^{\psi_1} \frac{e^{2\alpha R|r_s-r|\tan\psi}}{\cos\psi} d\psi + \int_0^{\psi_2} \frac{e^{-2\alpha R|r_s-r|\tan\psi}}{\cos\psi} d\psi_s \approx \sum_n g_n(r_s-r,z) + Y_n(r_s-r,z)$$

$$g_1(r_s-r,z) \approx e^{-2\alpha Rz} \frac{1}{2\alpha R|r_s-r|} [e^{2\alpha Rz}\cos\psi_1 - e^{-2\alpha R(l-z)}\cos\psi_2] \quad (A5.3)$$

$$Y_1(r_s-r,z) = e^{-2\alpha Rz} \frac{1}{2\alpha R|r_s-r|} \{\int_0^{\psi_1} \sin\psi e^{2\alpha R|r_s-r|\tan\psi} d\psi - \int_0^{\psi_2} \sin\psi e^{-2\alpha R|r_s-r|\tan\psi} d\psi\} \quad (A5.4)$$

$$g_2(r_s - r, z) \approx \frac{1}{(2\alpha R|r_s - r|)^2}[\cos^2\psi_1 \sin\psi_1 + \cos^2\psi_2 \sin\psi_2 e^{-2\alpha Rl}] \quad (A5.5)$$

$$Y_2(r_s - r, z) = -e^{-2\alpha Rz}\frac{1}{(2\alpha R|r_s - r|)^2}\{\int_0^{\psi_1} e^{2\alpha R|r_s - r|\tan\psi}\frac{d}{d\psi}[\cos^2\psi \sin\psi]d\psi$$
$$+ \int_0^{\psi_2} e^{-2\alpha R|r_s - r|\tan\psi}\frac{d}{d\psi}[\cos^2\psi \sin\psi)]d\psi\} \quad (A5.6)$$

Substituting these results into (A1.1) yields

$$E(r,z) = e^{-2\alpha Rz}\int_0^1 B(r_s)r_s\{\int_0^{\psi_1}\frac{e^{2\alpha R|r_s - r|\tan\psi}}{\cos\psi}d\psi dr_s + \int_0^{\psi_2}\frac{e^{-2\alpha R|r_s - r|\tan\psi}}{\cos\psi}d\psi\}dr_s$$
$$= \sum_n e^{-2\alpha Rz}\int_0^1 B_n(r_s - r, z)\{\sum_n g_n(r_s - r, z) + Y_n(r_s - r, z)\}r_s dr_s \quad (A5.7)$$

where

$$g_1(r_s - r, z) = \frac{1}{2\alpha R|r_s - r|}(\cos\psi_1 - e^{-2\alpha Rl}\cos\psi_2) \quad (A5.8)$$

$$g_2(r_s - r, z) = \frac{1}{(2\alpha R|r_s - r|)^2}[\cos^2\psi_1 \frac{d}{d\psi_1}(-\cos\psi_1) + \cos^2\psi_2 \frac{d}{d\psi_2}(-\cos\psi_2)e^{2\alpha Rl}] \quad (A5.9)$$

$$g_3(r_s - r, z) = -\frac{1}{(2\alpha R|r_s - r|)^3}\{[\cos^2\psi_1 \frac{d}{d\psi_1}\cos^2\psi_1 \frac{d}{d\psi_1}(-\cos\psi_1) -$$
$$-\cos^2\psi_2 \frac{d}{d\psi_2}\cos^2\psi_2 \frac{d}{d\psi_2}(-\cos\psi_2)e^{2\alpha Rl}\} \quad (A5.10)$$

$$g_4(r_s - r, z) = \frac{1}{(2\alpha R|r_s - r|)^4}e^{-2\alpha Rz}[(\cos^2\psi_1 \frac{d}{d\psi_1})^3(-\cos\psi_1)e^{2\alpha Rz}$$
$$+ (\cos^2\psi_2 \frac{d}{d\psi_2})^3(-\cos\psi_2)e^{-2\alpha R(l-z)}] \quad (A5.11)$$

…………………………………………………………………………………………..

$$g_n(r_s - r, z) = \frac{1}{(2\alpha R|r_s - r|)^n} e^{-2\alpha Rz} [(-\cos^2 \psi_1 \frac{d}{d\psi_1})^{n-1}(\cos \psi_1) e^{2\alpha Rz}$$

$$+ (-1)^n (-\cos^2 \psi_2 \frac{d}{d\psi_2})^{n-1}(\cos \psi_2) e^{2\alpha R(l-z)}]$$

(A5.12)

$$Y_n(r_s - r, z) = -\frac{1}{(2\alpha R|r_s - r|)^n} e^{-2\alpha Rz} \int_0^{\psi_1} e^{2\alpha R|r_s - r|\tan\psi} (-1)^n (-\frac{d}{d\psi}\cos^2 \psi)^{n-1}[\frac{d}{d\psi}(\cos\psi)]d\psi +$$

$$- \frac{1}{(2\alpha R|r_s - r|)^n} e^{-2\alpha Rz} \int_0^{\psi_2} e^{2\alpha R|r_s - r|\tan\psi} (-1)^n (-\frac{d}{d\psi}\cos^2 \psi)^{n-1}[\frac{d}{d\psi}(\cos\psi)]d\psi$$

where

$$\tan \psi_1 = \frac{z}{|r_s - r|}, \quad \tan \psi_2 = \frac{l-z}{|r_s - r|}, \quad \cos \psi_1 = \frac{1}{\sec \psi_1} = \frac{|r_s - r|}{\sqrt{|r_s - r|^2 + z^2}},$$

$$\sin \psi_1 = \sqrt{1 - \cos^2 \psi_1} = \sqrt{1 - \frac{|r_s - r|^2}{|r_s - r|^2 + z^2}} = \sqrt{\frac{z^2}{|r_s - r|^2 + z^2}}$$

(A5.13)

Substituting (A5.13) into (A5.8) (A5.11) yields

$$g_1(r_s - r, z) = \frac{1}{2\alpha R|r_s - r|} (\cos \psi_1 - e^{-2\alpha Rl} \cos \psi_2) = \frac{1}{2\alpha R|r_s - r|} \{\frac{|r_s - r|}{\sqrt{|r_s - r|^2 + z^2}}$$

$$- \frac{|r_s - r|}{\sqrt{|r_s - r|^2 + (l-z)^2}}\} = \frac{1}{2\alpha R} \{\frac{1}{\sqrt{|r_s - r|^2 + z^2}} - \frac{1}{\sqrt{|r_s - r|^2 + (l-z)^2}}\}$$

(A5.14)

$$g_2(r_s - r, z) = \frac{1}{(2\alpha R|r_s - r|)^2} [\cos^2 \psi_1 \frac{d}{d\psi_1}(-\cos \psi_1) + \cos^2 \psi_2 \frac{d}{d\psi_2}(-\cos \psi_2) e^{2\alpha Rl}]$$

$$= \frac{1}{(2\alpha R|r_s - r|)^2} \{\frac{z|r_s - r|^2}{(|r_s - r|^2 + z^2)^{\frac{3}{2}}} + \frac{(l-z)|r_s - r|^2}{[(|r_s - r|^2 + (l-z)^2]^{\frac{3}{2}}} e^{2\alpha Rl}\}$$

(A5.15)

$$= \frac{1}{(2\alpha R)^2} \{\frac{z}{(|r_s - r|^2 + z^2)^{\frac{3}{2}}} + \frac{(l-z)}{[(|r_s - r|^2 + (l-z)^2]^{\frac{3}{2}}}$$

$$g_3(r_s - r, z)$$
$$= -\frac{1}{(2\alpha R |r_s - r|)^3}\{[\cos^2\psi_1 \frac{d}{d\psi_1}\cos^2\psi_1 \frac{d}{d\psi_1}(-\cos\psi_1)$$
$$-\cos^2\psi_2 \frac{d}{d\psi_2}\cos^2\psi_2 \frac{d}{d\psi_2}(-\cos\psi_2)e^{2\alpha Rl}\}$$
$$= -\frac{1}{(2\alpha R |r_s - r|)^3}\{\cos^2\psi_1 \frac{d}{d\psi_1}(\cos^2\psi_1 \sin\psi_1) - e^{-2\alpha Rl}\cos^2\psi_2 \frac{d}{d\psi_2}(\cos^2\psi_2 \sin\psi_2)\}$$
$$= -\frac{1}{(2\alpha R |r_s - r|)^3}\{(\cos^5\psi_1 - 2\cos^3\psi_1 \sin^2\psi_1) - e^{-2\alpha Rl}(\cos^5\psi_2 - 2\cos^3\psi_2 \sin^2\psi_2)\}$$
$$= -\frac{1}{(2\alpha R |r_s - r|)^3}\{(3\cos^5\psi_1 - 2\cos^3\psi_1) - e^{-2\alpha Rl}(3\cos^5\psi_2 - 2\cos^3\psi_2)\}$$
$$= -\frac{1}{(2\alpha R |r_s - r|)^3}\{3(\frac{|r_s - r|}{\sqrt{|r_s - r|^2 + z^2}})^5 - 2(\frac{|r_s - r|}{\sqrt{|r_s - r|^2 + z^2}})^3\}$$
(A5.16)

$$g_4(r_s - r, z) = \frac{1}{(2\alpha R |r_s - r|)^4}e^{-2\alpha Rz}[(\cos^2\psi_1 \frac{d}{d\psi_1})^3(-\cos\psi_1)e^{2\alpha Rz}$$
$$+ (\cos^2\psi_2 \frac{d}{d\psi_2})^3(-\cos\psi_2)e^{-2\alpha R(l-z)}]$$
$$= \frac{1}{(2\alpha R)^4}\{\frac{-15|r_s - r|^2}{[|r_s - r|^2 + z^2]^3} + \frac{6}{[|r_s - r|^2 + z^2]^3}\}\frac{z}{\sqrt{|r_s - r|^2 + z^2}}$$
$$+ \frac{1}{(2\alpha R)^4}\{-\frac{15|r_s - r|^2}{[|r_s - r|^2 + (l-z)^2]^3}$$
$$+ \frac{6}{[|r_s - r|^2 + (l-z)^2]^2}\}\frac{l-z}{\sqrt{|r_s - r|^2 + (l-z)^2}}e^{-2\alpha Rl}]$$
(A5.17)

**Appendix 6. Numerical computation program (in maple)**

**1 For the initial energy density**

To calculate the radial distribution of energy density by changing the value of the radial coordinate x0 and other relevant variables

**N=1**

```
restart;x0:=0.5;
B :=(929/13440)^(-2)*add(10^7*(1/(3.16))*(Int(((3/32)*x^5-(343
/5760)*x^7+(1345/129024)*x^9-(109/184320)*x^11+(1/98304)*x^13)
*((432-z-9.535*k)/((x-x0)^2+(432-z-9.535*k)^2)^.5-(z+9.535*k)/
((x-x0)^2+(z+9.535*k)^2)^.5), x = 0 .. 1)), k = 0 .. 45);with(plots);
plot(B,z=0..432);
```

N=2
```
restart;x0:=0.5; B := 
(929/13440)^(-2)*10^14*(1/(3.16))^2*(Int(((3/32)*x^5-(343/5760)
*x^7+(1345/129024)*x^9-(109/184320)*x^11+(1/98304)*x^13)*((432
-z)/((x-x0)^2+(432-z)^2)^1.5+z/((x-x0)^2+z^2)^1.5), x = 0 .. 1));
with(plots); plot(B, z = 0 .. 432);
```

N=3
```
 restart;x0:=0.15;B := (929/13440)^(-2)*
(10^7/(3.16))^3*(Int(((3/32)*x^5-(343/5760)*x^7+(1345/129024)*
x^9-(109/184320)*x^11+(1/98304)*x^13)*((-(x-x0)^2+2*(z+0)^2)/(
(x-x0)^2+(z+0)^2)^2.5+((x-x0)^2-2*(432-z+0)^2)/((x-x0)^2+(432-
z+0)^2)^2.5), x = 0 .. 1)); with(plots); plot(B, z = 0 .. 432);
```

N=4
```
 restart;x0:=0.5; B :=
(929/13440)^(-2)*(10^7/(3.16))^4*(Int(((3/32)*x^5-(343/5760)*x
^7+(1345/129024)*x^9-(109/184320)*x^11+(1/98304)*x^13)*((6/((x
-x0)^2+(z+0*k)^2)^2-15*(x-x0)^2/((x-x0)^2+(z+0*k)^2)^3)*(z+0*k
)/((x-x0)^2+(z+0*k)^2)^.5+(6/((x-x0)^2+(432-z-0*k)^2)^2-15*(x-
x0)^2/((x-x0)^2+(432-z-0*k)^2)^3)*(432-z-0*k)/((x-x0)^2+(432-z
-0*k)^2)^.5), x = 0 .. 1)); with(plots); plot(B, z = 0 .. 432);
```

## 2. For the energy density of flowing fluid

N=1

```
 restart;x0:=0.5;B :=
(929/13440)^(-2)*add(10^7*(1/(3.16))*(Int(((3/32)*x^5-(343/576
0)*x^7+(1345/129024)*x^9-(109/184320)*x^11+(1/98304)*x^13)*((4
32-z-9.535*k)/((x-x0)^2+(432-z-9.535*k)^2)^.5-(z+9.535*k)/((x-
x0)^2+(z+9.535*k)^2)^.5), x = 0 .. 1)), k = 0 .. 45); with(plots);
plot(B, z = 0 .. 432);
```

N=2

```
restart; B := 
(929/13440)^(-2)*add((10^7/(3.16))^2*(Int(((3/32)*x^5-(343/576
0)*x^7+(1345/129024)*x^9-(109/184320)*x^11+(1/98304)*x^13)*(43
2-z-9.55*k)/((x-.7)^2+(432-z-9.55*k)^2)^1.5+(z+9.55*k)/((x-.7)
^2+(z+9.55*k)^2)^1.5, x = 0 .. 1)), k = 0 .. 45); with(plots);
logplot(B, z = 0 .. 432);
```

N=3

```
restart; x0:=0.15;B := 
(929/13440)^(-2)*add((10^7/(3.16))^3*(Int(((3/32)*x^5-(343/5760
)*x^7+(1345/129024)*x^9-(109/184320)*x^11+(1/98304)*x^13)*((-
(x-x0)^2+2*(z+9.532*k)^2)/((x-x0)^2+(z+9.532*k)^2)^2.5+((x-x0)
^2-2*(432-z-9.532*k)^2)/((x-x0)^2+(432-z-9.532*k)^2)^2.5), x = 
0 .. 1)), k = 0 .. 45); with(plots); logplot(B, z = 0 .. 432);
```

**N=4**

```
restart; x0:=.5; B:= 
(929/13440)^(-2)*add((10^7/(3.16))^4*(Int(((3/32)*x^5-(343/5760
)*x^7+(1345/129024)*x^9-(109/184320)*x^11+(1/98304)*x^13)*((6
/((x-x0)^2+(z+9.53*k)^2)^2-15*(x-x0)^2/((x-x0)^2+(z+9.53*k)^2)
^3)*(z+9.53*k)/((x-x0)^2+(z+9.53*k)^2)^.5+(6/((x-x0)^2+(432-z-
9.53*k)^2)^2-15*(x-x0)^2/((x-x0)^2+(432-z-9.53*k)^2)^3)*(432-z
-9.53*k)/((x-x0)^2+(432-z-9.53*k)^2)^.5), x = 0 .. 1)), k = 0 .. 
45); with(plots); logplot(B, z = 0..432);
```